\ifdefined\pdfoutput\pdfoutput=1\fi
\documentclass[11pt]{article}
\usepackage[a4paper,margin=1in]{geometry}
\usepackage[T1]{fontenc}
\usepackage{times}
\usepackage{booktabs}
\usepackage{graphicx}
\usepackage{amsmath,amssymb}
\usepackage{array}
\usepackage[font=small,labelfont=bf]{caption}
\usepackage{flafter}
\usepackage{url}
\usepackage[colorlinks=true,linkcolor=black,citecolor=black,urlcolor=blue]{hyperref}
\newcommand{\botrule}{\bottomrule}
\renewcommand{\footnotetext}[1]{\par\vspace{0.4em}\footnotesize\raggedright #1\par}
\begin{document}
\title{\bfseries Partial pooling beats synthetic respondents as estimators, not as covariates: a preregistered small-sample tournament}
\author{Howard Kim\\[0.3em]
\small College of AI Convergence, Seoul Cyber University, Seoul, South Korea\\
\small\url{howardkim@iscu.ac.kr}}
\date{}
\maketitle

\begin{abstract}
\noindent Survey estimates for population segments, such as media use by sex and age,
rest on few respondents per cell, and a growing literature offers synthetic
respondents generated by large language models to fill the gap. That literature has not tested them against the small-area estimators a
methodologist would otherwise use. We ran ten estimators, preregistered, on identical
household-clustered splits of a Korean media panel across five calibration
budgets and three holdout regimes. A multilevel partial-pooling model of the MRP
family led every calibration-only estimator at every budget, matching at one
hundred respondents what direct estimation reaches with roughly four times as
many. On unobserved segments, information fixed as a per-segment prior was inert
for synthetic panel and previous wave alike; attached as a bias correction it
transferred, at the price of estimation noise. No registered synthetic advantage
over MRP survived the paired decision rule. Entered instead as a covariate inside
the multilevel model, an exploratory hybrid, the same panel improved on MRP on
unobserved segments: the information survives where the registered constructions
did not. Interval coverage stayed below nominal throughout and worsened with
budget for partial pooling, contrary to prediction.

\medskip\noindent\textbf{Keywords:} small-area estimation, multilevel regression and poststratification, synthetic survey respondents, calibration weighting, preregistration, interval coverage
\end{abstract}

\section{Introduction}
Survey estimates are usually needed at a finer grain than the surveys producing
them were designed to support: a regulator wants adoption rates for men in their
twenties, a product team for women over seventy. Cross sex with seven age bands
and a nationally representative sample of eight thousand respondents becomes
fourteen cells, the smallest holding fewer than 250 people, where direct
estimation is noisiest.

The standard response is to borrow strength, from Stein's paradox (Efron \& Morris,
1975) through the area-level model that made it usable for official statistics
(Fay \& Herriot, 1979) to the modern synthesis (Rao \& Molina, 2015). Gelman and
Little (1997) recast poststratification as a hierarchical logistic regression,
letting the amount of pooling be estimated rather than assumed, and Park et al.
(2004) combined it with population cell counts to estimate state opinion from
national polls. The result, multilevel regression with poststratification (MRP),
is now the default for subnational estimates, and refinements continue (Gao et
al., 2021).

A newer and far less settled option has appeared alongside them: large language
models answering survey items in the voice of a synthetic respondent. Argyle et
al. (2023) showed that demographic backstories reproduce recognisable patterns of
political attitudes, and a rapid literature followed on whether such panels can
substitute for human samples (Ahnert et al., 2026; Bisbee et al., 2024), where
they fail (Dominguez-Olmedo et al., 2024; Santurkar et al., 2023), and which
response artefacts they carry (Braun, 2025; Tjuatja et al., 2024). If synthetic
panels carry usable signal about a segment, they could cover segments a real
survey cannot afford to reach, including segments no wave observed.

Attaching external information to a small survey is not itself new. One family
attaches the source as a prior and updates, as Wi\'sniowski et al. (2020) do with a
nonprobability sample as prior information for a smaller probability sample;
another learns the error where it is observable and transfers the correction, as
Marble and Clinton (2026) do in calibrating MRP against auxiliary quantities with
known margins; calibration weighting has advanced in parallel (Ben-Michael et al.,
2024). Our two registered synthetic estimators are instances of these families,
not new constructions.

Two things are missing from that picture. First, the synthetic-response
literature has begun to move from resemblance to use, but not yet to a direct
comparison with the estimators it would replace.
Huang et al. (2025) turn simulated responses into confidence sets for human
population quantities with an average-case coverage guarantee; Ye and
Yoganarasimhan (2026) ask how to allocate a fixed budget of human respondents
across questions when LLM predictions exist for all of them; Choi et al. (2026)
compare prompting, rectification and fine-tuning for aligning a simulator to a
small human pilot; Gordon et al. (2026) pass simulated respondents through the
same MRP pipeline as their human sample and find that post-stratification leaves
the residual distributional error where it was. None of these compares a synthetic panel with the small-area estimators a
survey methodologist would otherwise use, at the sample size they can afford,
for a segment the survey may not have reached.

Second, the two axes that decide practical value have not been crossed against a
partial-pooling baseline. The budget axis is not itself new, since Wi\'sniowski et
al. (2020) vary the probability sample from 50 to 1,000 respondents; what is
missing is its combination with segments the survey never reached and with a
synthetic panel judged against a standard small-area baseline. The small-area
literature has already warned that
its own methods are conditional: across eighty-nine issues, MRP's advantage
depended on strong predictors and adequate sample (Buttice \& Highton, 2013); our
conditioning axis is the one a practitioner controls, the calibration budget.

The question follows: when a small real calibration sample is combined with
external information, what survives once a standard small-area baseline is
present, and where does each approach stop working?

We run ten estimators against each other on identical household-clustered splits
of a national media panel, across five calibration budgets from 50 to 800
respondents and three holdout regimes, with the design, the estimators, the
metrics and the decision rule fixed in advance by preregistration (osf.io/w6s84):
direct estimation, raking, regression, MRP, fully Bayesian beta-binomial
shrinkage, previous-wave carry-forward, and a synthetic panel attached as a bias
correction and as a Bayesian prior. The contribution is a comparison, not a new
estimator, and it builds on our
earlier study (Kim \& Cho, 2026), which validated the same synthetic panel against
the same survey and already reported real-only regression and direct-estimation
baselines, a learning curve in the calibration fraction, cell holdout, cross-year
transfer of the correction, and the teen-cell composition mismatch, but lacked a
partial-pooling baseline. This paper adds the multilevel and Bayesian estimators,
fixes the budget grid, holdout regimes, metrics and decision rule in advance, and
re-evaluates the earlier conclusions on identical paired splits; Supplementary
Table S3 sets it beside the adjacent work on purpose, estimand, budget variation,
unobserved segments and baseline. Four findings follow.

A fixed per-segment prior is inert where there is no data to update it. On an
unobserved segment, an estimator that attaches external information as a
cell-specific prior mean returns that prior unchanged, its error flat in the
budget, for a synthetic panel and a previous wave alike; priors that pool across
cells keep learning.

Bias correction transfers, but it imports estimation noise. It pays only when
the source's bias is large relative to the noise of estimating it: at every
budget for a heavily biased synthetic panel, only past several hundred
respondents for an accurate previous wave.

A synthetic panel cannot represent a segment its persona source does not
contain. Our persona set contains no one under nineteen, so its ``teens'' cell is
composed entirely of nineteen-year-olds while the survey's spans ten to nineteen,
and its estimate matches real nineteen-year-olds far better than the cell it
stands for. The constraint is checkable before a study is designed; whether other
sources or models could reduce the error we did not test.

The panel's information survives inside the multilevel model. A hybrid
built after the tournament, the registered MRP with the synthetic cell rates as
a covariate, improves on MRP on unobserved segments at every budget for one
generating model and at the smallest budgets for the other. What lost to
partial pooling was not the synthetic information but the two ways of attaching
it that the field, and our registration, had used. Alongside these, MRP at one
hundred respondents matches direct estimation at roughly four hundred.

These findings share a methodological implication: neither the inertness of a
fixed per-segment prior nor the conditionality of correction is visible in an
evaluation confined to observed segments at a single sample size. We propose
unobserved segments and a budget range as a minimum standard for evaluating any proposal to attach external information
to survey estimation.

\section{Methods}
\subsection{Data}

The real data are the 2023, 2024 and 2025 waves of the Korea Media Panel Survey,
an annual household panel run by the Korea Information Society Development
Institute (KISDI), in which all eligible household members respond, so
observations are clustered within households and identifiers are available;
survey weights are supplied. After restricting to the seven age bands used for
segmentation, the 2024 analysis sample is 8,675 individuals in 4,016 households;
2025 has 8,395. The microdata are not redistributable under KISDI terms; access is
by application through the survey's official site.

The synthetic data are the public persona responses released with our earlier
study, generated from NVIDIA's Nemotron-Personas-Korea persona set. Two generating
models are included and analysed separately throughout, never averaged: Gemini and
EXAONE, each for 2024 and 2025. They enter this study only as cell-level aggregate
rates by sex and age band. Supplementary Table S4 gives the generating models,
generation period, persona allocation, prompt conditions and item sets; every
analysis is conditional on this fixed panel and none varies its generation.

Eight binary usage items are analysed: AI services, OTT, YouTube, short-form
video, SNS, messenger, metaverse, and paid content subscription.

\subsection{Estimand and comparison planes}

The estimand is the usage rate within a segment defined by the full cross of sex
(2) and age band (7), fourteen cells. The evaluation target for a cell is the
survey-weighted cell mean over the full sample of the relevant wave, 2024 for the
no-holdout and cell-holdout axes and 2025 for the temporal axis; fixing it
independently of the split makes the metrics comparable across splits, estimators
and regimes, and applies uniformly to a held-out cell.

The 2023 wave lacks the AI-service and short-form items, so the two previous-wave
estimators are undefined for them. The primary plane is the six items present in
all waves, the only one on which the full field competes; the secondary plane is
all eight items without those two estimators, a check that the ordering is not an
artefact of item selection. The planes are never pooled.

\subsection{Design}

Three factors are fully crossed.

\emph{Calibration budget}: n $\in$ \{50, 100, 200, 400, 800\} real respondents, drawn by
household: identifiers are shuffled and accumulated until the count of eligible
individuals first reaches the budget, so the realised size may exceed it by at
most one household.

Items are scored on complete responses, dropped per item and never imputed, so
the YouTube and short-form items have a smaller effective sample (7,346 of 8,675
in 2024, 7,196 of 8,395 in 2025), as registered.

\emph{Holdout regime}: none; cell holdout, one sex-by-age cell designated unobserved
and scored alone, each of the fourteen in turn; and temporal holdout, calibrating
on 2024 and scoring against 2025. A cell-holdout replicate combines fourteen
separately drawn calibration samples, one per removed cell, and pools the
fourteen held-out predictions into one cell MAE, so it is a set of
cross-validation folds rather than a single sample; pairing across estimators is
preserved, but a split-stability share on this axis is a share of fold sets, not
the probability of winning on one segment.

\emph{Estimator}: ten estimators, the two synthetic-source ones split by generating
model, twelve reported variants.

Split randomisation reinitialises a fixed seed at each budget, which makes each
budget's calibration sample a subset of the next one's, contrary to what the
registration assumed; every estimator is still evaluated on the same 200 splits
within a condition, which is what the paired comparisons require, and we reran
the whole tournament with independent draws (Deviations).

\subsection{Estimators}

Table 1 names the estimators and their sources; Supplementary Note S2 gives
priors, hyperpriors, concentration estimation and empty-cell fallbacks.

\begin{table}[htbp]
\caption{Estimator specifications.}\label{table1}%
\footnotesize
\begin{tabular}{@{}l>{\raggedright\arraybackslash}p{0.26\textwidth}>{\raggedright\arraybackslash}p{0.22\textwidth}cc@{}}
\toprule
Code & Estimator & Information used & Reports intervals & Items defined \\
\midrule
(a) & Direct weighted cell estimation & Calibration set & No & 8 \\
(b) & Raking to sex, age band and education margins & Calibration set + frame margins & No & 8 \\
(c) & Age-and-sex weighted linear regression & Calibration set & No & 8 \\
(d) & MRP (multilevel logistic with partial pooling) & Calibration set & Yes & 8 \\
(e) & Beta-binomial shrinkage (fully Bayesian) & Calibration set & Yes & 8 \\
(f) & Previous-wave carry-forward & Previous full wave, no calibration set: 2023, or 2024 under temporal holdout & No & 6 \\
(g) & Previous wave as prior, Bayesian updating & Calibration set + 2023 wave as prior, on every axis & No (point only) & 6 \\
(h) & Synthetic panel with signature bias correction & Calibration set + synthetic (2 models) & No & 8 \\
(i) & Synthetic panel as prior, Bayesian updating & Calibration set + synthetic (2 models) & Yes & 8 \\
(j) & Grand mean (uninformative floor) & Calibration set & No & 8 \\
\botrule
\end{tabular}
\footnotetext{Specifications were fixed in the preregistration before the tournament was run. Estimators (f) and (g) are defined on the six items the 2023 wave also carries; the remaining estimators are defined on all eight. Raking includes an education margin because raking on the cell-defining margins alone is algebraically identical to (a). Estimator (f) does not consume the calibration budget, an information advantage quantified in the Results.}
\end{table}
 Sources
shift with the axis. Without holdout and under cell holdout, (f) and (g) draw on
the 2023 wave and (h) and (i) on the 2024 synthetic panel. Under temporal
holdout, calibrating on 2024 against 2025 targets, (f) carries the full 2024 wave
forward, (g) keeps its 2023 prior and updates it with the 2024 sample, and the
synthetic estimators are applied to the 2025 panel, (h) learning its correction
on the 2024 panel. The axis is a retrospective cross-wave transfer test, not a
forward forecast: the 2025 synthetic responses were generated in 2026 by models
served in mid-2026 (Gemini's stated knowledge cutoff is January 2025, EXAONE's is
undisclosed), so information later than the calibration year could in principle
have reached them. What the axis isolates is whether an estimator fitted on 2024
data transfers to 2025 targets. The distinction that organises our results is
how an estimator attaches external information: as a prior to update with
the calibration data, (g) centring a beta-binomial prior on the previous wave's
cell rate and (i) on the synthetic panel's; as a bias correction, (h)
regressing the difference between synthetic and calibration rates on age across
the observed cells and subtracting the fitted bias from every cell; or,
unregistered and described below as a post-hoc hybrid, as a covariate inside
the multilevel model.

The baselines begin with (a) direct weighted cell estimation and (b) raking, the
classical alternative to modelling the outcome: it adjusts the weights by
iterative proportional fitting (Deming \& Stephan, 1940), later placed in a
general calibration framework by Deville and S\"arndal (1992). We rake to the
full-wave margins for sex, age band and collapsed education. The education margin
is not optional: raking on the cell-defining margins alone makes the raking
factors constant within each cell and the estimate algebraically identical to
(a), an identity we verify empirically; at the smallest budgets raking sometimes
cannot satisfy its margins at all (Supplementary Note S1). The remaining
baselines are (c) an age-and-sex weighted regression on cell rates; (d) a
multilevel logistic partial-pooling model with a non-centred age random effect,
registered as MRP, the family's usual name, in which the poststratification step
is the identity because the estimand is the cell itself; (e) a fully Bayesian
beta-binomial; (f) previous-wave carry-forward, which consumes no calibration budget, an information advantage quantified in
the Results with a budget-matched variant; and (j) the calibration grand mean, the
uninformative floor.

\emph{Metrics and coverage plane}: cell MAE, RMSE and worst-cell error are reported on
the six-item plane unless noted. Interval coverage, computed for the three
interval-reporting estimators by averaging cell-level coverage indicators within
a replicate, was stored only over all eight items, so every coverage figure in
this paper, including Figure 3 and the H7 verdict, is on the eight-item plane; we
label the difference wherever it matters.

\subsection{Weighting convention}

The evaluation target is weighted. Estimators (a), (b), (c), (f), (h) and (j)
compute weighted cell means from the calibration sample; (d), (e), (g) and (i) are
fitted to unweighted cell counts, because at these sample sizes a weighted
likelihood distorts the effective sample size and miscalibrates the posterior.
Those four therefore target the unweighted cell proportion, which differs from
the weighted target by 1.98 points on average and 13.30 at the extreme across the
2024 cells and items. Scoring the whole no-holdout axis against both definitions
(Supplementary Note S3) leaves MRP first among calibration-only estimators at
every budget, moves the synthetic-versus-MRP comparison against the synthetic
side under the unweighted target, and changes ranks only within adjacent pairs
the decision rule cannot separate. Coverage is reported against both targets. We
do not use pseudo-likelihood devices (Parker et al., 2023); whether weighting
helps or hurts precision depends on how the weights relate to the outcome (Little
\& Vartivarian, 2005).

\subsection{Decision rule}

No hypothesis is tested by a parametric significance test. Within a budget,
holdout regime and plane, each estimator yields a per-split cell MAE over the 200
shared splits; for a pair we report the mean per-split difference with its 2.5th
and 97.5th percentiles. One estimator outperforms another only when that interval
excludes zero, and the difference is practically meaningful only when its mean
also reaches 1.0 percentage point, our prespecified smallest effect of interest;
differences that clear the first bar but not the second are reported as
distinguishable but negligible.

These are percentile intervals over repeated calibration draws from a fixed
dataset, not confidence intervals for a population parameter, and they answer a
narrower question than ``distinguishable'' suggests: an interval that excludes zero
means the ranking held on at least 97.5 per cent of individual draws, a central
95 per cent interval lying wholly on one side of zero, which we call
\emph{split-stable}. Whether the \emph{average} difference is separated from zero is a
different question, answered by the Monte Carlo standard error of the mean, and
``not split-stable'' is neither equivalence nor evidence that the mean difference
is zero. The two disagree in 22 of the 108 registered contrasts, where the mean is
located away from zero with high precision while the sign flips on a minority of
draws; we keep the registered rule as primary, report the mean with its Monte
Carlo error alongside it wherever they diverge, and give the central contrast in
all three readings in Supplementary Table S1.

Two properties of the rule should be read alongside it. We apply no familywise
correction, as registered: multiplicity is controlled structurally, confirmatory
language being confined to a contrast set fixed in advance (sized at 88, run as
108 because the registration requires (h) and (i) to be reported separately for
each generating model), with every other comparison reported but labelled
exploratory and the full matrix published. And pairing does not cancel noise
equally: (f) ignores the calibration sample, so its paired difference inherits
only the other estimator's variability, and (g) and, less so, (i) behave the same
way under cell holdout, where an unobserved cell leaves their posterior mean at
the prior mean.

Every results table includes the uninformative floor (j); an estimator that fails
to beat it is marked as failing.

\subsection{Preregistration and prior access to the data}

The design, estimator specifications, metrics and decision rule above were
registered before the tournament was run (osf.io/w6s84, registered 24 August
2026, embargoed until the analyses were complete and public since). The data
existed and had been analysed before this study. Our earlier work
(Kim \& Cho, 2026) used the same KISDI extract and produced the synthetic panel,
and before registering we had run a single-budget pilot at n = 100 without
holdout covering all estimators, a diagnostic to decide how coverage should be
scored, and simulation validation of every self-implemented estimator. Our
directional predictions were therefore informed by the pilot; we treat them as
prespecified predictions rather than confirmatory tests, draw primary evidence
from the four remaining budgets and both holdout axes, and mark every table cell
that re-presents the pilot design point.

\subsection{Implementation validation}

Each self-implemented estimator was validated against known ground truth in
simulation before being admitted, with pass criteria fixed in advance. The
validation changed four implementations: an empirical-Bayes beta-binomial was
replaced by a fully Bayesian version after its intervals collapsed to a coverage
of 0.26 in one scenario, two samplers were reparameterised after boundary-rate
items produced heavy divergence, and the raking estimator's fixed-iteration
scheme was replaced after simulation exposed non-convergence at small samples and
a faulty early-exit test.

Posteriors were sampled with the No-U-Turn sampler, two chains of 1,000 tuning
and 1,000 retained draws, target acceptance 0.9. The 200 replicates of a
condition were fitted as one batched model, 2,560 batched fits in the registered
run, every one completing with zero divergent transitions; the post-hoc
strengthened variant also had none, and the post-hoc hybrid's 27 are reported
with it. Convergence statistics were not computed during the production run, so we
recomputed them afterwards on conditions chosen to bracket the claims (the
smallest and largest budgets, each without and with a cell held out; the temporal
axis shares the no-holdout sampling geometry), on the three items nearest to and
furthest from the probability boundary, for every interval-reporting estimator,
the strengthened MRP and both hybrids. Across 66 batched fits and 253,800
parameters the largest R-hat was 1.019 and 169 parameters exceeded 1.01, seventy
in the strengthened variant and seventy-five in the hybrids; the smallest bulk
and tail effective sample sizes were 224 and 92, in the EXAONE hybrid with a cell
held out at n = 800 on the rarest item, below the conventional 400; and three
transitions diverged, in the Gemini hybrid at n = 800. Comparing each cell's
2.5th and 97.5th posterior percentiles across the two chains, the median gap is
0.3 points at n = 50 and 0.1 at n = 800, but the largest is 25 points, in cells
the smallest calibration draws leave without respondents; chain-specific coverage differed by a median of 0.002 and a maximum of 0.025,
so some condition-specific second-decimal coverage values are sensitive to the
chain, and individual endpoints in near-empty cells more so. We changed nothing as a result.

\subsection{Robustness of the ranking}

The ranking survives the checks registered for it: RMSE tracks the MAE ordering
closely while worst-cell error does not, the one systematic divergence being the
contradiction described in the Results; the eight-item plane moves ranks without
changing a verdict of consequence; and individual rather than household splitting
at n = 100 moves the largest mean error by 0.29 points, consistent with the design
effect of 0.99 measured before registration (Supplementary Note S1).

\subsection{A stronger multilevel variant}

The registered MRP is a basic member of its class, with exchangeable rather than
ordered age effects and no sex-by-age interaction, so a synthetic advantage over
it could reflect a weak baseline. To test this we built a stronger variant after
the tournament, with a BYM2 age mixture
and a sex-by-age interaction (Supplementary Note S2), validated it by simulation
before touching the tournament data, and reran all fifteen conditions on the same 200 splits with zero divergences; every comparison involving it is
exploratory. On
observed segments it changes nothing of consequence, never gaining a point over
the registered MRP and separating from it only at n = 50, so the registered MRP is
not a weak baseline where the main rankings live; on unobserved segments the
improvement is larger (11.09 to 9.86 points at n = 50), and there the variant
matters.

\subsection{A hybrid: the synthetic panel inside the multilevel model}

Neither registered synthetic construction has a multilevel structure, so a loss
to MRP on unobserved segments could mean that the synthetic information is
worthless there or that the way it was attached is. To
separate them we built, after the tournament, a hybrid that keeps the registered
MRP intact and adds the synthetic panel's cell rate as a covariate on the logit
scale, $\gamma$(\emph{s}$_j$ $-$ mean \emph{s}), with \emph{s}$_j$ the logit of the synthetic rate for cell \emph{j}
and $\gamma$ $\sim$ Normal(0, 0.5) (Supplementary Note S2). At $\gamma$ = 0 it is the registered
MRP; on an unobserved cell \emph{s}$_j$ is still available, so the panel enters through a
coefficient learned on the observed cells rather than as a fixed prior mean. Its prior scale was set by simulation before touching the tournament data
(Supplementary Note S2); it was then run on the same 200 splits for both
generating models and all fifteen conditions, and is likewise exploratory.

\subsection{Deviations from the preregistration}

We list every departure from the registered plan; Supplementary Note S4 gives
each in full. (1) \emph{Budgets are nested rather than independent}; an
independent-seed rerun of the entire tournament moves the mean paired difference
by a median of 0.04 points across the 118 registered contrast cells and changes
two verdicts, both on the rule's boundary, and the registered run is primary.
(2) \emph{The registration's stated basis for 200 splits was wrong}, conflating the
Monte Carlo error of a percentile estimate with the width of the split-difference
distribution; we kept 200 splits and the rule and correct the justification.
(3) \emph{H4 has 40 contrasts, not the 20 the registration sized}, because the same
registration requires the two generating models to be reported separately.
(4) \emph{Previous-wave source under temporal holdout}: the registration's estimator
table names the 2023 wave for (f) and (g) while its baseline rule adds (f) to the
temporal axis as an out-of-time baseline; the implementation follows the rule
for (f), which uses the 2024 wave there, and the table for (g), both fixed in
code before the run. (5) \emph{A post-hoc twelve-cell analysis} is reported beside the
registered result, not in its place. (6) \emph{Convergence statistics were computed after the run} and changed nothing.
(7) \emph{Three registered exploratory analyses were not run}: alternative raking
margins, the shrinkage parameter's trajectory
in the budget, and the reversed temporal axis; the full pairwise matrix is
published for mean absolute error only.

\subsection{Use of AI tools}

Large language models were used in two roles: as the object of study, they
generated the synthetic persona responses analysed here (Gemini and EXAONE); as a
writing and coding aid, an AI assistant helped implement the estimator code,
run the pipeline, and draft this manuscript. All analysis decisions, the
preregistration and the interpretation are the author's; every reported number
was produced by the archived code, and digital object identifiers are recorded
only where verified.

\section{Results}
Unless noted, figures are for the primary six-item plane and errors are in
percentage points, abbreviated to points. Intervals are percentile intervals over
the 200 shared calibration draws, not population confidence intervals. Cells
reporting the n = 100 no-holdout design point re-present the pilot that informed
our predictions and are marked in the tables.

\subsection{Ranking and cost equivalence}

Figure 1 and Supplementary Table S7 give cell MAE by budget and holdout regime. Under no holdout,
raking tracks direct estimation to within 0.05 points at every budget, confirming
the algebraic near-identity described in Methods, while MRP sits below both
throughout, falling from 8.76 to 3.02 points as the budget grows from 50 to 800.

\begin{figure}[htbp]
\centering
\includegraphics[width=\textwidth]{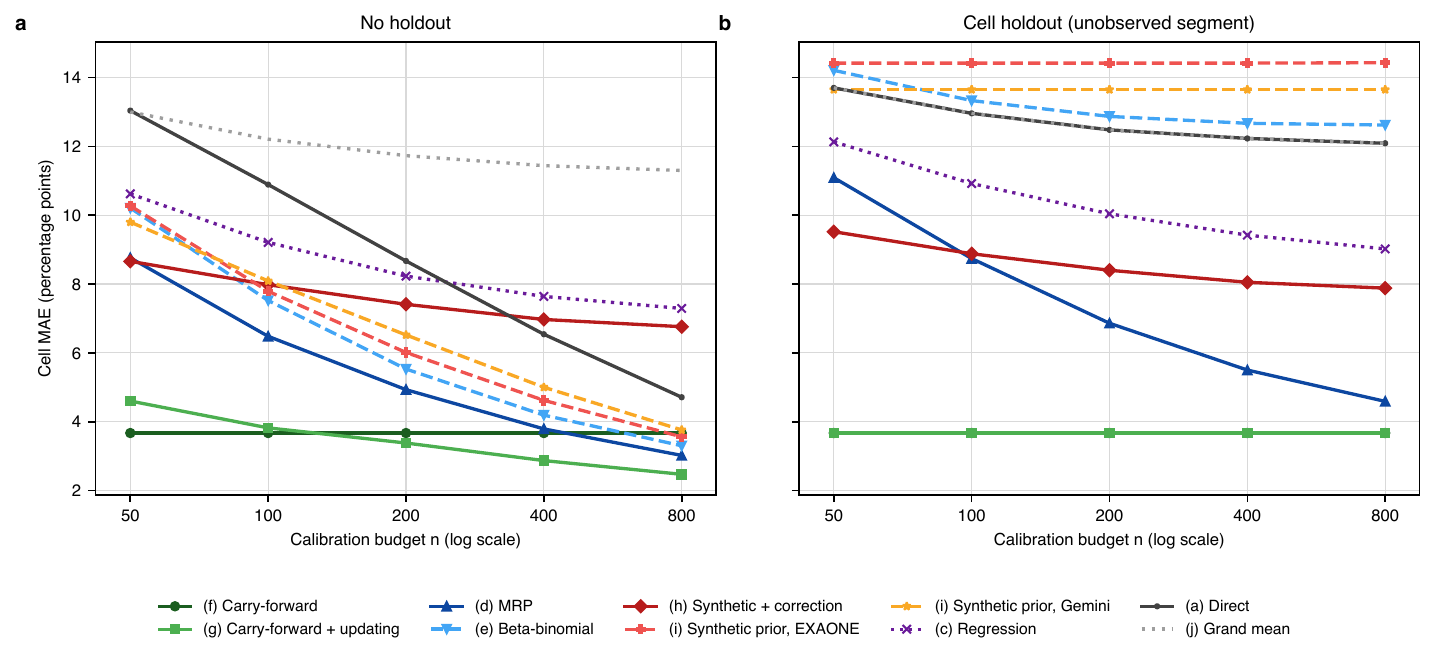}
\caption{\textbf{Learning curves by calibration budget and holdout regime.} Cell mean absolute error in percentage points, averaged over 200 shared household-clustered calibration draws, primary six-item plane. Raking (b) and the Gemini-based correction are omitted for legibility; (h) is EXAONE; Supplementary Table S7 gives every value. (\textbf{a}) No holdout. (\textbf{b}) Cell holdout, the target cell absent from the calibration sample; (a) coincides with (j) exactly and (f) with (g) to within 0.003 points. Log budget axis.}\label{fig1}
\end{figure}

Our registered prediction was that direct estimation would catch up with MRP
between 200 and 400 respondents. It does not: the paired (a) minus (d) difference
narrows from 4.28 to 1.69 points but its interval excludes zero at every budget,
including 800 ([0.82, 2.61]). Within the tested range, partial pooling has no
crossover point.

To express the ranking in sample sizes we fitted the direct estimator's learning
curve
(MAE = 57.7 $\times$ $n^{-0.367}$, R$^2$ = 0.982) and inverted it to express each
estimator's accuracy as the budget direct estimation would need to match it
(Supplementary Table S2). At one hundred respondents MRP matches what direct
estimation reaches near 384, an equivalent budget roughly four times larger. That
is a sample-size equivalence, not a measured cost saving: a point on a curve
fitted to five aggregate means, assuming cost linear in respondents and counting
no generation, fieldwork or analysis cost; the previous-wave figures extrapolate beyond n = 800 and are not claimed. The direction of the
multiplier separates the estimators. MRP's rises from 3.4 at n = 50 to a plateau
near 4, while the synthetic estimator's falls monotonically from 3.5 to 0.4,
dropping below one by n = 400, where it is worse than buying more respondents; the
age-sex regression follows the same descent and the beta-binomial, like MRP,
plateaus. Partial pooling does not compound without limit, its multiplier peaking
at n = 400, but it keeps its advantage where estimators with saturating
functional forms lose theirs to residual bias; whether it would itself saturate
at larger budgets this design cannot say.

Previous-wave carry-forward sits at 3.67 points regardless of budget, ahead of
every other estimator at the two smallest budgets and overtaken by the shrinkage
combination from n = 200. But it consumes no calibration budget, using the whole
previous wave, 9,699 respondents, against its competitors' 50 to 800. Drawing the
same budget from that wave by household makes the comparison like-for-like:
matched on budget, the previous wave loses to MRP at every budget (5.63
against 3.02 at n = 800), and a same-year sample of the same size gives 4.71, so
the one-year gap costs under a point. The registered finding that previous-wave
estimators beat every calibration-only estimator thus reframes as a large free
sample beating a small purchased one, fair as a decision problem and not as a
methods comparison; the budget-matched analysis is exploratory.

\subsection{Updating versus correcting}

The clearest structural result appears under cell holdout, where the target
segment contributes nothing to the calibration sample. Estimator (i), which
attaches the synthetic panel as a prior and updates, returns 14.42 points at
n = 50 and 14.43 at n = 800 (Supplementary Table S8); the flatness is the
result. With no
observations in the held-out cell the posterior mean stays at the prior and the
estimator reports the raw synthetic rate, worse than the uninformative floor of
12.09 at n = 800.

The flatness is structural. Writing the beta-binomial prior for cell \emph{j} as
Beta(p$_{0j}\kappa$, (1$-$p$_{0j}$)$\kappa$) and observing \emph{y$_j$} users among \emph{n$_j$} respondents, the
posterior mean is (\emph{y$_j$} + p$_{0j}\kappa$)/(\emph{n$_j$} + $\kappa$); an unobserved cell has
\emph{y$_j$} = \emph{n$_j$} = 0, so the posterior mean collapses to p$_{0j}$ exactly, for any $\kappa$ and
any amount of data elsewhere. The posterior itself does not collapse, since $\kappa$ is
shared across cells and estimated from the observed ones, so the unobserved
cell's spread, p$_{0j}$(1 $-$ p$_{0j}$)/($\kappa$ + 1) given $\kappa$, still moves with the rest of the
data; what is fixed is the point estimate.
What is inert is therefore the point estimate of a conjugate update onto a
fixed, cell-specific prior mean, not Bayesian priors in general: priors that
pool the mean behave differently, (e) falling from 14.21 to 12.62 under cell
holdout as the budget grows and MRP, sharing through the age random effect, from
11.09 to 4.59. The same mechanism appears with a different source: (g), which
attaches the \emph{previous wave} as a prior and updates, is identical to plain
carry-forward (f) on this axis to within 0.003 points across all 1,000 splits, a
residue of prior clipping. On an unobserved segment, updating a fixed
per-segment prior mean leaves the point estimate where it started.

Bias correction behaves differently. Estimator (h), which learns the same
panel's bias on the observed cells and transfers the correction, reaches 9.52
points at n = 50, nearly five points better than the updating form on identical
information. We expected this to generalise, and it does not: a previous-wave analogue of (h)
makes things worse at n = 100, moving 3.67 to 5.60, and helps
only by n = 800, at 3.27, because the correction must be estimated from the
calibration sample and imports that noise along with the structure. What separates the two cases is the
size of the bias: the synthetic panel is heavily biased, so there is much to
remove; the previous wave is already accurate, so noise dominates until the
budget is large. The condition can be stated: attach external information as a correction rather
than a prior when the segment may be unobserved, but expect it to pay only when
the source's bias exceeds the noise of estimating it and transfers in the form
the correction assumes, here a linear function of age. Two
sources on one dataset illustrate the condition; they do not establish a general
threshold.

\subsection{Where the synthetic niche is}

Our central registered hypothesis was that synthetic-source estimators would beat
model extrapolation on unobserved segments. As registered, across the
fourteen-cell design, it fails: of forty contrasts, ten favour the synthetic side,
eight are indistinguishable, and twenty-two favour MRP or the beta-binomial
(Table 2).

\begin{table}[htbp]
\caption{Verdicts on the registered predictions.}\label{table3}%
\footnotesize
\begin{tabular}{@{}l>{\raggedright\arraybackslash}p{0.30\textwidth}c>{\raggedright\arraybackslash}p{0.30\textwidth}@{}}
\toprule
Hypothesis & Registered prediction & Contrasts & Verdict \\
\midrule
H1 & Direct estimation crosses MRP between n = 200 and 400 & 5 & Failed --- MRP leads at every budget; no crossover \\
H2 & MRP leads all calibration-only estimators at n $\leq$ 200 & 12 & Partly supported --- 11 of 12; the registered claim was universal \\
H3 & Previous-wave estimators beat all calibration-only estimators at n $\leq$ 200 & 36 & Supported --- 36 of 36, but see the budget-matched comparison \\
H4 & Synthetic estimators beat (d) and (e) on unobserved segments & 40 & Partly failed --- 10 of 40 favour synthetic, 22 favour the baselines \\
H5 & EXAONE-based variants beat Gemini-based variants on MAE & 10 & Failed on accuracy --- all 10 contrasts indistinguishable \\
H6 & (g) overtakes (f) at n $\geq$ 400 & 5 & Partly supported --- reversal distinguishable at n = 400, practically meaningful only at 800 \\
H7 & Coverage falls below nominal, worst at the smallest budgets & 12 & Partly supported --- below nominal everywhere (all eight items); budget clause holds in 6 of 12 \\
\botrule
\end{tabular}
\footnotetext{Contrast counts are the prespecified comparisons for each prediction. A prediction is supported only when the paired split-percentile interval excludes zero in the predicted direction. The registration sized the H4 set at 20 contrasts; the run produced 40 because estimators (h) and (i) are reported separately for each generating model, as the same registration requires. H7 has two clauses and only the first holds throughout. Deviations are listed in full in the Methods.}
\end{table}

The composition matters more than the total. Against the
beta-binomial, which has no covariate structure and can only shrink an unobserved
cell toward a global mean, the synthetic estimator wins at every budget by four
to five points. Against MRP, which extrapolates through the age-sex
structure, it leads only at n = 50, and there our two criteria diverge: the mean
difference is $-$1.56 points with a Monte Carlo standard error of 0.08 and MRP
loses on 92.5 per cent of draws (of replicate fold sets, not of single segments),
but the interval [$-$3.79, 0.47] includes zero, so the registered rule records no
advantage. The mean favours MRP from n = 200, but the registered interval
excludes zero only from n = 400 (Figure 2). That result stands as primary,
though it is computed on a design containing two cells the synthetic panel
cannot represent.

\begin{figure}[htbp]
\centering
\includegraphics[width=\textwidth]{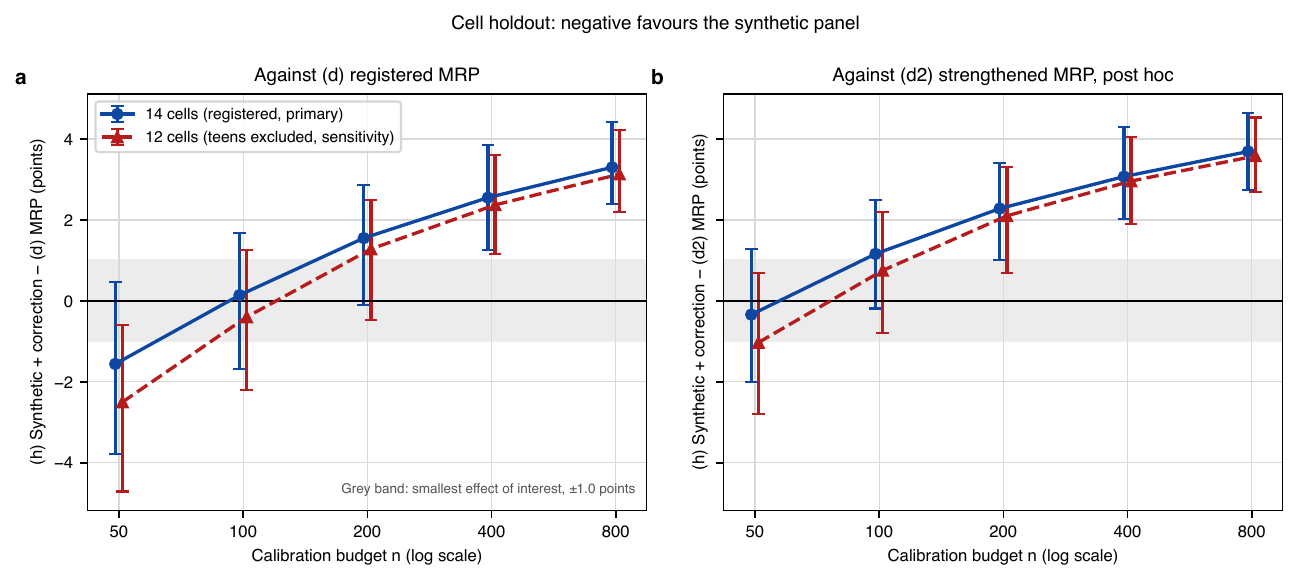}
\caption{\textbf{Where the synthetic panel's advantage on unobserved segments begins and ends.} Paired per-split difference in cell mean absolute error between the corrected synthetic estimator (h, EXAONE) and a multilevel model under cell holdout, in points; negative favours the synthetic panel. (\textbf{a}) Against the registered MRP (d). (\textbf{b}) Against the strengthened variant (d2), post hoc; the one interval that excludes zero in the synthetic panel's favour in \textbf{a} (12 cells, n = 50) widens back over it. Points are means over 200 shared draws, whiskers the 2.5th to 97.5th percentiles; circles with solid lines are the registered fourteen-cell design, triangles with dashed lines the twelve-cell sensitivity analysis excluding the two teen cells outside the persona source's population. Grey band: smallest effect of interest, $\pm$1.0 points. Series are offset horizontally; log budget axis.}\label{fig2}
\end{figure}

\subsection{The persona population boundary}

The persona set underlying the synthetic panel contains no one younger than
nineteen, so its teens cell consists entirely of nineteen-year-olds, while the
survey's teens cell spans ten to nineteen, of whom only 15.5 per cent are
nineteen.

Table 3 tests that explanation: the synthetic estimate differs from real
nineteen-year-olds by 7.45 points, from the full teens cell by 11.84, and from
those eighteen and under by 12.89, against internal heterogeneity of 11.79
points; the real nineteen-year-olds number only 75, so the comparison supports
the ordering of the gaps rather than their values, and it is descriptive rather
than a paired test.

\begin{table}[htbp]
\caption{The synthetic teens estimate matches real nineteen-year-olds, not the cell it represents.}\label{table5}%
\footnotesize
\begin{tabular}{@{}>{\raggedright\arraybackslash}p{0.44\textwidth}rr@{}}
\toprule
Compared with & Mean absolute difference (points) & n \\
\midrule
Real respondents aged 19 & 7.45 & 75 \\
Real respondents aged 18 and under & 12.89 & 409 \\
The full teens cell (ages 10-19) & 11.84 & 484 \\
Aged 19 versus aged 18 and under (within the survey) & 11.79 & 484 \\
\botrule
\end{tabular}
\footnotetext{Mean absolute difference between the EXAONE synthetic estimate for the teens cell and three real comparison groups, averaged over the primary six-item plane and both sexes, 2024 wave. The persona source contains nobody younger than nineteen, so the synthetic teens cell consists entirely of nineteen-year-olds while the survey's teens cell spans ages ten to nineteen, of whom 15.5 per cent are nineteen. The final row gives the survey cell's own internal heterogeneity. With only 75 real nineteen-year-olds the 7.45 figure carries appreciable noise; the ordering of the gaps, not the precise value, is what the comparison supports. The comparison is descriptive, not a paired test.}
\end{table}
 The panel's error in that cell, 12.3 points without holdout at
n = 100 and 13.1 under cell holdout, is a mismatch of population definition
rather than a failure of prediction: a synthetic persona panel cannot represent a
segment outside its source's population. We did not test whether other prompts
or models close this gap, but neither adds ages the source does not contain, so
the remedy is to align segment boundaries with the source's population or
exclude the cells that cannot be aligned.

We therefore recomputed the cell-holdout comparison excluding the two teen cells
(Supplementary Table S1). At n = 50 the synthetic estimator against MRP moves
from $-$1.56 [$-$3.79, 0.47], not distinguishable, to $-$2.50 [$-$4.71, $-$0.59], which
excludes zero and exceeds the one-point threshold; at n = 100 the point estimate
crosses zero (+0.14 to $-$0.40) but the interval still contains it, and from
n = 400 MRP leads under both definitions.

This subset favours the synthetic panel, since the teen cells cost it more
than they cost MRP (13.1 against 9.7 at n = 100). Three considerations justify reporting it: the removal criterion is population
definition, not performance; the mismatch was documented by our earlier study
(Kim \& Cho, 2026) before these results existed; and it applies symmetrically to
every estimator. The result is a
candidate niche, not an established one: a synthetic advantage at n = 50 on
unobserved segments the persona source can represent, gone by n = 100 and
reversed from n = 400, resting on a post-hoc subset of one dataset.

The candidate does not survive a stronger baseline. The strengthened MRP variant
(Methods) never separates from the registered model by as much as a point on
observed segments, so nothing above changes; on held-out segments, where
structure helps most, the n = 50 point difference reverses sign, the n = 200
contrast becomes a clear MRP lead, and on the twelve-cell subset the one interval
that had excluded zero in the synthetic panel's favour, $-$2.50 [$-$4.71, $-$0.59],
widens back over it, $-$1.03 [$-$2.79, +0.69]. Both estimators in that comparison are post hoc, so this is a sensitivity
analysis rather than a verdict: the candidate niche survives the registered MRP
and not an ordinary strengthening of it.

\subsection{The synthetic panel as a covariate}

The two registered constructions leave one question open: whether the synthetic
information itself lost to partial pooling, or only the way it was attached. The
hybrid described in the Methods gives the panel the same multilevel structure as
MRP and lets the data set its weight; Supplementary Tables S5 and S6 give every
contrast, and three results matter.

First, the panel adds accuracy where the registered constructions could not.
Against the registered MRP the Gemini-based hybrid is split-stable under cell
holdout at every budget, by 2.75 points at n = 50 falling to 0.93 at n = 800,
above the one-point threshold through n = 400, and at n = 50 on the other two
axes ($-$1.67 without holdout, $-$1.69 temporal). The EXAONE-based hybrid gains 0.83
and 0.50 points under cell holdout at n = 50 and 100 and 0.52 on the temporal
axis at n = 50, split-stable but below the threshold, and nothing distinguishable
elsewhere. Neither is ever split-stably worse than the registered MRP.

Second, information and structure are not interchangeable. Against the
strengthened MRP, which adds structure without information, the EXAONE-based
hybrid is indistinguishable except for one contrast below the threshold, while
the Gemini-based hybrid stays ahead under cell holdout through n = 400 ($-$1.53 at
n = 50, $-$1.07 at n = 100). At n = 50 under cell holdout the ordering is Gemini hybrid 8.33, corrected
EXAONE panel 9.52, strengthened MRP 9.86, EXAONE hybrid 10.26 and registered MRP
11.09 points.

Third, the better estimator is not the better covariate. As a corrected
estimator the Gemini panel is worse than the EXAONE panel on every axis, and as a
fixed prior its coverage is the worst in the study; its raw cell rates are only
slightly closer to the target (13.7 against 14.4 points of error), too little to
explain a two-point gap at n = 50. As a covariate it is worth more because the
hybrid uses the panel's between-cell pattern on the logit scale, rescaled by a
coefficient and offset by the sex and age effects, not its levels. Both hybrids
beat the correction from n = 200 on every axis (by 1.8 to 6.4 points) and the
fixed prior under cell holdout at every budget (by 4 to 10); their intervals
inherit MRP's coverage shortfall (0.80 to 0.89 without holdout), and 27 of 2,560,000 retained transitions diverged, all at n = 800 and 23 of them
in items outside the primary plane. Both sides of these comparisons are post hoc, but their bearing on the
registered question is direct: the panel's registered advantage did not survive
the baselines, but its information does, once attached inside the model the
baselines use rather than beside it.

\subsection{Uncertainty}

Three estimators report intervals. None reaches its nominal level against the
full-wave benchmark (Figure 3); coverage in this section is over all eight items
(Methods). Against the weighted target and without holdout, MRP's realised 95 per cent
coverage runs from 0.858 at n = 50 to 0.819 at n = 800. The registration
predicted a shortfall, which holds everywhere, and placed its maximum at the
smallest budgets, which holds in six of twelve estimator-by-holdout combinations
and reverses in six: it holds for the beta-binomial and both synthetic priors without
holdout and under cell holdout (the Gemini-based prior's shortfall falls from
0.328 to 0.163), and reverses for MRP on all three axes and for every estimator on
the temporal axis, where the shortfall roughly doubles (MRP 0.264 to 0.427).
Where it reverses, more data narrows the intervals while the misspecification
bias stays put; where it holds, the prior is wide enough at small budgets that
shrinking it is a net gain. We assess Bayesian intervals by frequentist coverage,
the calibrated-Bayes programme (Little, 2006), and, as registered, did not adjust
the models to reach nominal coverage.

\begin{figure}[htbp]
\centering
\includegraphics[width=\textwidth]{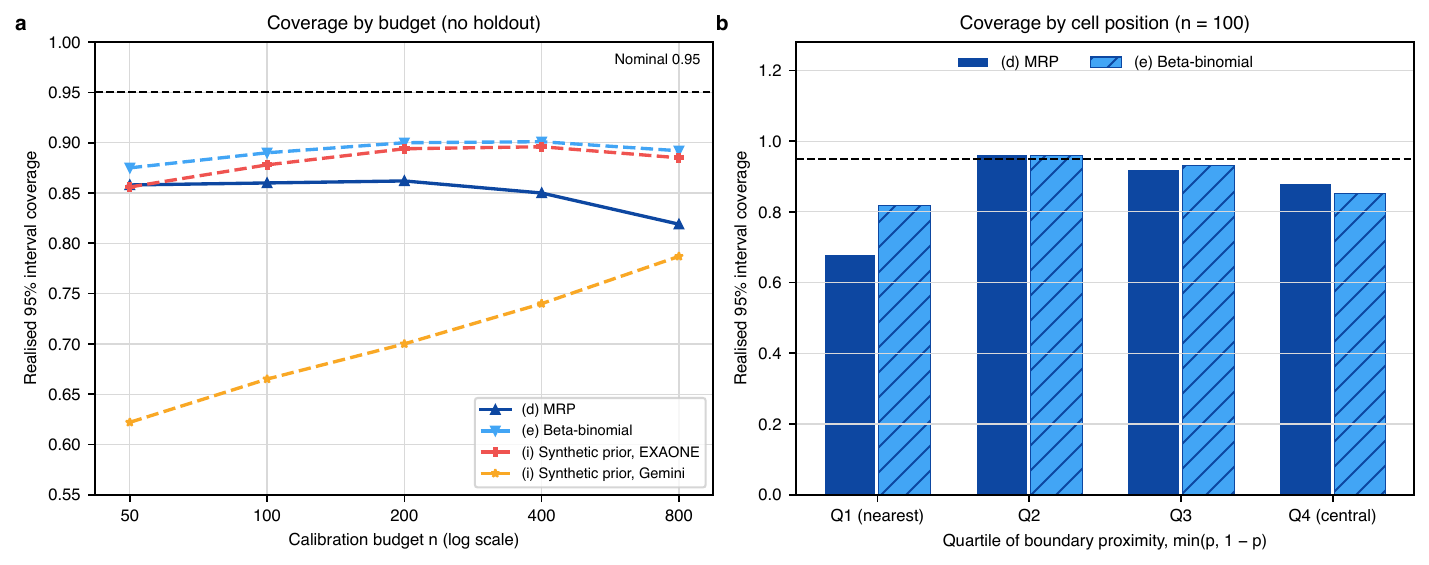}
\caption{\textbf{Realised coverage of nominal 95 per cent intervals}, computed over all eight items (Methods). (\textbf{a}) Coverage against the weighted target without holdout, by budget, for the three interval-reporting estimators, Gemini- and EXAONE-based synthetic priors separate; the dashed line marks the nominal level. (\textbf{b}) Coverage at n = 100 by the cell's proximity to the probability boundary, min(p, 1 $-$ p), in quartiles from nearest (Q1) to most central (Q4).}\label{fig3}
\end{figure}

Where the shortfall occurs is informative. In the quartile of cells nearest the
probability boundary, coverage is 0.680 for MRP and 0.819 for the beta-binomial,
against 0.92 to 0.96 in the middle quartiles; the relationship is not monotone,
so we claim only that coverage collapses at boundary cells, a localised failure
consistent with the smoothing assumption rather than a design-effect artefact
(the measured design effect is 0.99), though it does not rule out the
weighted-target mismatch described in the Methods and we do not claim to have
identified the cause.

Model dependence is small in point accuracy and large in the intervals. Under the
registered rule the prediction that EXAONE-based variants beat Gemini-based ones
on MAE fails, no contrast being split-stable, although EXAONE leads the corrected
estimator at every budget by 0.30 to 0.51 points, each more than six Monte Carlo
standard errors from zero but below the one-point threshold. Coverage is where
the difference is large: the Gemini-based synthetic prior runs 0.62
to 0.79 without holdout and 0.53 to 0.56 under cell holdout, against 0.85 to 0.91
for EXAONE. Choice of generating model moves the point estimates a little and
the intervals a great deal.

\subsection{Worst-cell error selects a different estimator}

We registered that all three metrics, and any contradiction among their
rankings, would be reported. One contradiction is systematic. Under cell holdout
the synthetic-prior estimator (i) ranks last of twelve on mean absolute error at
every budget, yet on worst-cell error it ranks third at n = 50 and n = 100, ahead
of MRP and of the corrected synthetic estimator: its worst cell is off by 42.4
points where MRP's is off by 49.5 and direct estimation's by 66.5. The mechanism
is the one that makes it a poor average estimator: anchored to a prior it cannot
leave, its worst-cell error is set by the prior's own bias rather than by
sampling variation, and in these data that bias was smaller than MRP's worst
extrapolation error, an observed ordering rather than a bound, since a prior
wrong by fifty points would produce a worst-cell error of fifty points. Mean and worst-case error therefore select different estimators.

\subsection{Registered hypotheses}

Table 2 gives the verdicts: one of the seven predictions supported outright, two
failed (H1 and H5 on accuracy) and four hold in part. H3 holds in all thirty-six
contrasts, previous-wave estimators beating every calibration-only estimator at
n $\leq$ 200, but the budget-matched comparison shows why that is not evidence about
the previous wave as a \emph{source}. Two exploratory analyses did not survive:
per-item testing does not support the impression that the synthetic panel beats
MRP on the newer items (at n = 100 no item separates; at n = 800 MRP leads on five
and the synthetic panel on none), and an ensemble prior averaging the two
generating models improves the corrected estimator slightly without reaching the
threshold and reverses at n = 800 for the prior-based form.

\section{Discussion}
The results support a compact rule for a practitioner. If a full previous wave
already exists, use it; adding a small current sample through a shrinkage
combination is worth it only from about 400 respondents. If no previous wave
exists and the segments are observed, use MRP: per unit of calibration budget it
is the strongest estimator we tested, and its advantage holds as the budget grows
while the alternatives' decays. If a segment is unobserved, the most accurate estimator we found at any budget
was the registered MRP with the synthetic panel's cell rates as a covariate, a
hybrid built after the tournament and therefore exploratory; the registered
bias-correction form beat the registered MRP only at fifty respondents, on a
post-hoc subset, and not against the strengthened baseline, so it is a signal to
replicate rather than a recommendation. Whatever the choice, do not take the intervals at face value: realised coverage
ran from 0.52 to 0.91 against a nominal 0.95, worst where width matters most.

Two implications follow. How the panel is attached matters more than which
panel it is: the same information, fixed as a per-segment prior, was inert on
unobserved segments; attached as a bias correction it transferred at the cost of
estimation noise; entered
as a covariate inside the multilevel model it improved on MRP itself. And align
segment boundaries with
the persona source's population before designing the study. Our persona set began at nineteen, our
segmentation crossed that boundary, and no reweighting of the personas on hand
could repair the resulting cell; the boundary is checkable in advance from the source's documentation. Whether
extrapolation from other sources or models could reduce the error we did not
test; the limit belongs to the source as sampled.

This study was designed to test whether a niche reported by our own earlier work
(Kim \& Cho, 2026) withstands standard baselines. Part of it does not: at n = 100
without holdout synthetic estimators do not beat MRP. Part of it may: on
unobserved segments at the smallest budget, for segments the persona source
covers, the corrected synthetic estimator held an advantage over the registered
MRP, on a post-hoc subset of one dataset and not against an ordinary
strengthening of the baseline. What the earlier study could not establish, for
want of a structural baseline, is how narrow that region is; our contribution is
a boundary and a location to test, not a refutation.

Beneath the specific comparisons lies a distinction that we think generalises
beyond this dataset. External information can enter a small-sample estimator in
at least three ways, and they behave differently: fixed to a segment as its prior
mean, it is inert where it would be most valuable; applied as a correction, it
transfers, but brings estimation noise proportional to how little bias there was
to remove; entered as a covariate inside a partial-pooling model, it is rescaled
by a coefficient the observed segments estimate, so it can help even when its
levels are wrong, which is why the panel that made the worse corrected estimator
was the more useful covariate. The first failure is escapable, since a prior that
pools across segments keeps learning, and it does not extend to constructions
such as Wi\'sniowski et al. (2020), whose prior sits on regression coefficients that
pool across cells by design. Neither failure is visible in an evaluation confined
to observed segments at a single sample size.

Several limitations bound these conclusions. Every result rests on a single
national panel, fourteen cells and eight binary items; the mechanisms should
transfer but the crossover locations are quantities of this dataset. Our preregistration missed the persona population boundary, so the
central registered hypothesis was tested on two cells the synthetic panel cannot
represent. Per-item intervals run about fourteen points wide at n = 100, so the
rejected item-level hypothesis is undetectable here rather than shown false. The temporal axis confounds
year-over-year change with panel attrition and refreshment, and because the 2025
synthetic responses were generated after 2025 it is a retrospective transfer
test, not a forecast. Cell holdout removes one cell at a time, so the other sex at
the same age stays observed; removing an entire age band, or several cells at
once, is a harder test we did not run. The twelve-cell recomputation and the
independent-seed rerun are sensitivity analyses on the same data, not
replications in a new population.

Our registered multilevel model uses an additive logit with an independent age
random effect; structured priors that encode age ordering reduce bias in MRP
settings (Gao et al., 2021), multilevel calibration weighting strengthens our
raking baseline (Ben-Michael et al., 2024), and pseudo-likelihood devices
reconcile a weighted target with an unweighted likelihood (Parker et al., 2023).
Beyond the strengthened variant and the hybrid, both post hoc, we use none of
them, so part of the coverage shortfall may be specific to these
basic forms; the hybrid's coverage is no better than MRP's. On the estimand mismatch we measured the exposure: scoring the whole no-holdout axis
against the unweighted target leaves MRP first among calibration-only estimators
at every budget and moves the synthetic comparison against the synthetic side,
so the target we report is the more generous of the two for that panel, which
makes our synthetic-versus-MRP comparisons conservative but not the coverage or
crossover findings, and both may be properties of this model rather than of
partial pooling. Finally, all reported intervals summarise variation across
repeated calibration draws from a fixed dataset rather than uncertainty about
the Korean population.

We therefore end with a proposal about evaluation, not about estimation. Any
claim that an external source improves survey estimation should be tested on at
least two axes: segments the source must extrapolate to, and a range of
calibration budgets. Observed segments at a single sample size cannot separate an estimator that
borrows structure from one that echoes its input. The standard costs little to meet,
and would have caught, in our own earlier work, the boundary we found only
here.

\section{Statements and Declarations}
\subsection{Ethical considerations}
All methods were carried out in accordance with relevant guidelines and
regulations. This study is a retrospective secondary analysis of de-identified
microdata; it involved no contact with human participants and no new data
collection. The Institutional Review Board of Sungkyunkwan University reviewed the
parent project, which analyses the same dataset, and determined it exempt from
review (approval number SKKU 2026-07-021, notified 9 July 2026), on the ground
that a retrospective analysis of already-collected secondary data poses no risk
to participants. The data were collected by KISDI for the Korea Media Panel
Survey, waves 14 to 16 (2023 to 2025); the analysis sample includes minors, 4.7
per cent of the 2024 records being aged eighteen or under and 5.6 per cent in the
ten-to-nineteen band, and all reported quantities are cell-level aggregates.

\subsection{Consent to participate}
Not applicable. The study analyses de-identified secondary microdata and recruited
no participants. The Institutional Review Board that granted the exemption above
waived the requirement for additional informed consent for this secondary
analysis. Recruitment and consent procedures for the original survey were
conducted by the Korea Information Society Development Institute under that
survey's own protocols.

\subsection{Consent for publication}
Not applicable. No data from any identifiable individual are reported. All
quantities reported are cell-level aggregates, and no individual-level records are
reproduced or redistributed.

\subsection{Declaration of conflicting interest}
The author declared no potential conflicts of interest with respect to the
research, authorship, and/or publication of this article.

\subsection{Funding statement}
The author received no financial support for the research, authorship, and/or
publication of this article.

\subsection{Data availability}
The reproduction package (archived as v1.2 at
\url{https://doi.org/10.5281/zenodo.22793037}) contains the analysis code, a single
entry point that runs the whole pipeline in order, the per-split metrics
underlying every interval reported here, and the aggregate tables and figures,
but neither input dataset. The Korea Media Panel Survey microdata are not
redistributable under KISDI terms of use; access is by application through the
survey's official portal (\url{https://stat.kisdi.re.kr}), after which the analysis
extract can be reconstructed with the recoding rules in the reproducibility
package of Kim and Cho (2026) (archived as v1.0 at
\url{https://doi.org/10.5281/zenodo.21397425}), which also holds the synthetic persona
responses. The preregistration is publicly available at \url{https://osf.io/w6s84}.
Analyses were run under Python 3.11.3 with PyMC 5.28.5, PyTensor 2.38.3, ArviZ
0.23.4, NumPy 2.4.6, pandas 3.0.3, SciPy 1.17.1 and Matplotlib 3.11.1; the
package pins these versions.

\section{References}
\begingroup
\setlength{\parindent}{0pt}
\setlength{\leftskip}{0.5in}
\setlength{\parskip}{0pt}
\hspace*{-0.5in}Ahnert, G., Haensch, A.-C., Plank, B., \& Strohmaier, M. (2026). Survey response generation: Generating closed-ended survey responses in-silico with large language models. In \emph{Proceedings of the 64th Annual Meeting of the Association for Computational Linguistics (Volume 1: Long Papers)} (pp. 41554--41577). \url{https://doi.org/10.18653/v1/2026.acl-long.1927}

\hspace*{-0.5in}Argyle, L. P., Busby, E. C., Fulda, N., Gubler, J. R., Rytting, C., \& Wingate, D. (2023). Out of one, many: Using language models to simulate human samples. \emph{Political Analysis}, \emph{31}(3), 337--351. \url{https://doi.org/10.1017/pan.2023.2}

\hspace*{-0.5in}Ben-Michael, E., Feller, A., \& Hartman, E. (2024). Multilevel calibration weighting for survey data. \emph{Political Analysis}, \emph{32}(1), 65--83. \url{https://doi.org/10.1017/pan.2023.9}

\hspace*{-0.5in}Bisbee, J., Clinton, J. D., Dorff, C., Kenkel, B., \& Larson, J. M. (2024). Synthetic replacements for human survey data? The perils of large language models. \emph{Political Analysis}, \emph{32}(4), 401--416. \url{https://doi.org/10.1017/pan.2024.5}

\hspace*{-0.5in}Braun, D. (2025). Acquiescence bias in large language models. In \emph{Findings of the Association for Computational Linguistics: EMNLP 2025} (pp. 11341--11355). \url{https://doi.org/10.18653/v1/2025.findings-emnlp.607}

\hspace*{-0.5in}Buttice, M. K., \& Highton, B. (2013). How does multilevel regression and poststratification perform with conventional national surveys? \emph{Political Analysis}, \emph{21}(4), 449--467. \url{https://doi.org/10.1093/pan/mpt017}

\hspace*{-0.5in}Choi, E. C., Kim, Y., Pugalenthi, P., Chen, H.-E., \& Huang, B.-R. (2026). Beyond the mean: Three-axis fidelity for aligning LLM-based survey simulators from small pilot data. \emph{Pluralistic Alignment Workshop, ICML 2026}. arXiv. \url{https://doi.org/10.48550/arXiv.2606.28963}

\hspace*{-0.5in}Deming, W. E., \& Stephan, F. F. (1940). On a least squares adjustment of a sampled frequency table when the expected marginal totals are known. \emph{The Annals of Mathematical Statistics}, \emph{11}(4), 427--444. \url{https://doi.org/10.1214/aoms/1177731829}

\hspace*{-0.5in}Deville, J.-C., \& S\"arndal, C.-E. (1992). Calibration estimators in survey sampling. \emph{Journal of the American Statistical Association}, \emph{87}(418), 376--382. \url{https://doi.org/10.2307/2290268}

\hspace*{-0.5in}Dominguez-Olmedo, R., Hardt, M., \& Mendler-D\"unner, C. (2024). Questioning the survey responses of large language models. In \emph{Advances in Neural Information Processing Systems} (Vol. 37). \url{https://doi.org/10.52202/079017-1458}

\hspace*{-0.5in}Efron, B., \& Morris, C. (1975). Data analysis using Stein's estimator and its generalizations. \emph{Journal of the American Statistical Association}, \emph{70}(350), 311--319. \url{https://doi.org/10.1080/01621459.1975.10479864}

\hspace*{-0.5in}Fay, R. E., \& Herriot, R. A. (1979). Estimates of income for small places: An application of James--Stein procedures to census data. \emph{Journal of the American Statistical Association}, \emph{74}(366a), 269--277. \url{https://doi.org/10.1080/01621459.1979.10482505}

\hspace*{-0.5in}Gao, Y., Kennedy, L., Simpson, D., \& Gelman, A. (2021). Improving multilevel regression and poststratification with structured priors. \emph{Bayesian Analysis}, \emph{16}(3), 719--744. \url{https://doi.org/10.1214/20-BA1223}

\hspace*{-0.5in}Gelman, A., \& Little, T. C. (1997). Poststratification into many categories using hierarchical logistic regression. \emph{Survey Methodology}, \emph{23}(2), 127--135.

\hspace*{-0.5in}Gordon, A., Petrova, N., Burden, J., Bosch, O. J., \& Ding, N. (2026). \emph{Do richer personas improve LLM survey simulation? A fidelity paradox} (SSRN Working Paper No. 7306183). SSRN. \url{https://ssrn.com/abstract=7306183}

\hspace*{-0.5in}Huang, C., Wu, Y., \& Wang, K. (2025). Uncertainty quantification for LLM-based survey simulations. In \emph{Proceedings of the 42nd International Conference on Machine Learning} (PMLR Vol. 267, pp. 25947--25971). \url{https://proceedings.mlr.press/v267/huang25am.html}

\hspace*{-0.5in}Kim, H., \& Cho, K. T. (2026). \emph{Distributional validity and calibration of a Korean synthetic persona panel for digital and AI service use: A secondary-data validation against the Korea Media Panel Survey} (arXiv:2608.28615). arXiv. \url{https://doi.org/10.48550/arXiv.2608.28615}

\hspace*{-0.5in}Kolenikov, S. (2014). Calibrating survey data using iterative proportional fitting (raking). \emph{The Stata Journal}, \emph{14}(1), 22--59. \url{https://doi.org/10.1177/1536867X1401400104}

\hspace*{-0.5in}Little, R. J. A. (2006). Calibrated Bayes: A Bayes/frequentist roadmap. \emph{The American Statistician}, \emph{60}(3), 213--223. \url{https://doi.org/10.1198/000313006X117837}

\hspace*{-0.5in}Little, R. J. A., \& Vartivarian, S. (2005). Does weighting for nonresponse increase the variance of survey means? \emph{Survey Methodology}, \emph{31}(2), 161--168.

\hspace*{-0.5in}Marble, W., \& Clinton, J. D. (2026). Improving small-area estimates of public opinion by calibrating to known population quantities. \emph{Political Analysis}, 1--14. \url{https://doi.org/10.1017/pan.2026.10044}

\hspace*{-0.5in}Park, D. K., Gelman, A., \& Bafumi, J. (2004). Bayesian multilevel estimation with poststratification: State-level estimates from national polls. \emph{Political Analysis}, \emph{12}(4), 375--385. \url{https://doi.org/10.1093/pan/mph024}

\hspace*{-0.5in}Parker, P. A., Janicki, R., \& Holan, S. H. (2023). A comprehensive overview of unit-level modeling of survey data for small area estimation under informative sampling. \emph{Journal of Survey Statistics and Methodology}, \emph{11}(4), 829--857. \url{https://doi.org/10.1093/jssam/smad020}

\hspace*{-0.5in}Rao, J. N. K., \& Molina, I. (2015). \emph{Small area estimation} (2nd ed.). Wiley. \url{https://doi.org/10.1002/9781118735855}

\hspace*{-0.5in}Santurkar, S., Durmus, E., Ladhak, F., Lee, C., Liang, P., \& Hashimoto, T. (2023). Whose opinions do language models reflect? In \emph{Proceedings of the 40th International Conference on Machine Learning} (PMLR Vol. 202, pp. 29971--30004).

\hspace*{-0.5in}Tjuatja, L., Chen, V., Wu, S. T., Talwalkar, A., \& Neubig, G. (2024). Do LLMs exhibit human-like response biases? A case study in survey design. \emph{Transactions of the Association for Computational Linguistics}, \emph{12}, 1011--1026. \url{https://doi.org/10.1162/tacl_a_00685}

\hspace*{-0.5in}Wi\'sniowski, A., Sakshaug, J. W., Perez Ruiz, D. A., \& Blom, A. G. (2020). Integrating probability and nonprobability samples for survey inference. \emph{Journal of Survey Statistics and Methodology}, \emph{8}(1), 120--147. \url{https://doi.org/10.1093/jssam/smz051}

\hspace*{-0.5in}Ye, Z., \& Yoganarasimhan, H. (2026). Rectification difficulty and optimal sample allocation in LLM-augmented surveys. arXiv. \url{https://doi.org/10.48550/arXiv.2604.17267}
\par\endgroup

\clearpage
\begingroup\setlength{\parskip}{6pt}\setlength{\parindent}{0pt}
\begin{center}{\Large\bfseries Supplementary Information}\end{center}

\textbf{Supplementary Note S1. Robustness of the ranking.}

RMSE tracks the MAE ordering closely (median rank correlation 0.97 across the
fifteen conditions, no estimator moving more than five places). Worst-cell error
does not, and the Results describe the one systematic divergence; the median
correlation there is 0.88 and the largest displacement is nine places. Read alone, worst-cell error separates the estimators much as MAE does at the
extremes: direct estimation's worst cell reaches 78.2 points at n = 50 and is
still 27.1 at n = 800, while MRP's falls from 38.9 to 19.5. The age-sex
regression is the exception, barely improving with budget (52.7 to 38.7),
consistent with a misspecification more data cannot fix.

The eight-item plane changes the ordering more than we expected but no verdict
of consequence. Ranks differ from the common-six plane in fourteen of fifteen
conditions (median rank correlation 0.98), almost all of it between estimators
the decision rule cannot separate. Of the 394 pairs
distinguishable on both planes, five reverse sign, and all five are the same
contrast, the EXAONE against the Gemini synthetic prior under cell holdout, where
the common-six difference is 0.77 points, below our smallest effect of interest,
and both estimators sit below the uninformative floor. We also reran the no-holdout axis at n = 100 with individual rather than
household splitting, as registered. Nothing of consequence moves: the largest
change in mean error is 0.29 points, every paired interval covers zero, and the
only rank change is an adjacent swap the decision rule cannot separate. This matches the design effect of
0.99 measured before registration and explains it: household members are
distributed across different sex-by-age cells, so within-cell clustering is
negligible even though the sampling is clustered. Carry-forward is identical under both schemes to the last
digit, as it must be, since it never touches the calibration sample.

Raking's small-sample limit. Raking with an auxiliary education margin failed to satisfy its margins in 13.5 per
cent of splits at n = 50 without holdout and 24.9 per cent under cell holdout,
falling to about 1 per cent at n = 100 and effectively zero from n = 200. This is a
structural property, not an implementation fault: at small samples the
sex-by-age-by-education cross-classification is sparse enough that no weight set
satisfies all margins simultaneously (Kolenikov, 2014). We retain these splits,
report the rate, and note that raking was in any case indistinguishable from direct
estimation throughout.

\textbf{Supplementary Note S2. Estimator specifications.}

Cell probabilities are written $p_j$ for the fourteen sex-by-age cells, with $y_j$
users among $n_j$ calibration respondents and $p_{0j}$ the external source's rate
for that cell.

(a) Direct estimation returns the survey-weighted cell mean of the calibration
sample; an empty cell falls back to the calibration grand mean, which makes (a)
and (j) identical under cell holdout. (b) Raking runs iterative proportional
fitting on sex, age band and education margins, then forms the same weighted cell
mean; the education margin is what separates it from (a), since raking on the
cell-defining margins alone is algebraically identical to direct estimation.
(c) fits a weighted linear regression of cell rates on sex and a linear age index
and predicts every cell from it.

(d) is a multilevel logistic model, $\mathrm{logit}(p_j) = \alpha + \beta\,\mathrm{sex}_j + u_{\mathrm{age}(j)}$,
with $u$ a non-centred normal random effect over the seven age bands,
$\alpha \sim \mathrm{Normal}(0, 1.5)$, $\beta \sim \mathrm{Normal}(0, 1)$ and the age scale
$\mathrm{HalfNormal}(1)$. The strengthened variant replaces the independent age
effect with a BYM2 mixture of a first-order random walk and an unstructured term,
the total scale keeping the same prior and the mixing proportion given
$\mathrm{Beta}(1, 1)$, and adds a sex-by-age interaction with scale
$\mathrm{HalfNormal}(0.5)$. (e) is a beta-binomial with
$p_j \sim \mathrm{Beta}(\mu_e \kappa_e, (1 - \mu_e)\kappa_e)$,
$\mu_e \sim \mathrm{Beta}(2, 2)$ and $\kappa_e \sim \mathrm{Gamma}(2, 0.1)$, both
hyperparameters estimated from all cells at once.

(f) carries a previous wave's weighted cell rate forward unchanged and consumes no
calibration budget. The wave is the most recent complete one before the target:
2023 without holdout and under cell holdout, 2024 under temporal holdout. (g)
centres a beta-binomial on the 2023 wave's rate on every axis,
$p_j \sim \mathrm{Beta}(p_{0j}\kappa, (1 - p_{0j})\kappa)$, and fixes a single
$\kappa$ shared by all cells by maximising the marginal likelihood; the posterior
mean is $(y_j + p_{0j}\kappa)/(n_j + \kappa)$. Under temporal holdout (g) thus
updates a 2023 prior with the 2024 calibration sample and is scored against 2025,
while (f) carries the full 2024 wave. (h) regresses the synthetic panel's bias on
the observed cells against a linear age index and subtracts the fitted bias from
the synthetic rate, falling back to the mean observed bias when the regression is
not identified; the bias is learned on the 2024 panel, the calibration year, and
under temporal holdout is subtracted from the 2025 panel. (i) uses the same prior
form as (g) with $p_{0j}$ from the synthetic panel for the target year, 2024 or
2025, but samples $\kappa \sim \mathrm{Gamma}(2, 0.1)$ by Hamiltonian Monte Carlo,
one $\kappa$ per split shared across cells, with $p$ marginalised analytically and
its posterior recovered by conditional composition; this is the only synthetic
estimator reporting intervals. (j) returns the calibration grand mean for every
cell.

The post-hoc hybrid (d+s) is (d) with one added term, $\gamma\,(s_j - \bar{s})$,
where $s_j$ is the logit of the clipped synthetic cell rate for the panel of the
target year and $\gamma \sim \mathrm{Normal}(0, 0.5)$; all other priors are those
of (d). The prior scale was set in simulation before any real data were used: with
$\gamma \sim \mathrm{Normal}(0, 1)$ an uninformative synthetic source cost 0.6 to
0.7 points at $n \leq 100$ under a sex-by-age interaction truth, above the
0.5-point gate, and 0.5 brought it to 0.06 to 0.31 points. Under the final prior
and a monotone age-curve truth, an exact or age-linearly biased source gains 0.6
to 4.7 points on observed cells and 4 to 15 points on a cell held out of the
calibration sample, with coverage between 0.94 and 0.98; under the interaction
truth the same sources gain $-$0.03 to 1.25 points with coverage between 0.915 and
0.975. Batched and single fits agree to within 0.5 points. Eight divergent
transitions occurred in three of the 64 batched fits, all at n = 800 under the
age-curve truth (one, five and two, of 2,000 transitions per fit); the gate
allowed one per cent per fit. The gate had seven criteria (no harm from an exact
source; harm from a biased, warped or unrelated source bounded at 0.5 points;
coverage at least 0.90; no out-of-range estimates and divergences at most one per
cent of transitions per fit; batch-single agreement; and a held-out-cell gain of
at least one point from an exact source); the simulation record is in the
reproduction package.

All rates are clipped to [0.001, 0.999] before use as prior centres. Sampling uses
two chains of 1,000 draws after 1,000 tuning steps at a target acceptance of 0.9.

\textbf{Supplementary Note S3. Scoring against the weighted and the unweighted target.}

The evaluation target is weighted. Estimators (a), (b), (c), (f), (h) and (j) also
compute weighted cell means from the calibration sample. Estimators (d), (e), (g)
and (i) are fitted to unweighted cell counts.

This is deliberate. At these sample sizes a weighted likelihood distorts the
effective sample size: weighting the counts makes the binomial denominator no
longer a count of observations, and the posterior is badly calibrated as a result.
Because the estimand is defined at the cell level and the model conditions on the
cell, the poststratification step is the identity.

The cost is that these four estimators target the unweighted cell proportion while the
evaluation target is the weighted cell mean, so any within-cell association between the
weights and the outcome appears as bias. This mismatch is not small. Across the eight
items and fourteen cells of the 2024 wave the two definitions differ by 1.98 points on
average and by 13.30 points at the extreme, which is two thirds of MRP's best observed
error, so it could in principle reorder the field rather than shift coverage alone.

We therefore scored the whole no-holdout axis against both definitions. The ranking is
stable where it matters. MRP is the best calibration-only estimator at every budget
under both targets, and its margin over the runner-up is if anything wider against the
unweighted target (1.61 against 1.45 points at n = 50). The comparison between the
corrected synthetic estimator and MRP moves against the synthetic side at every budget
(at n = 800, +3.74 becomes +4.08), so the weighted target on which we report is the
more generous of the two for the synthetic panel. Rank changes do occur, but they are
confined to adjacent pairs the decision rule cannot separate: at most three places, and
none at n = 200. The estimators built on unweighted counts gain most from the switch,
as expected, the synthetic prior by about 0.5 to 0.6 points and MRP by 0.25. The one
large movement in the other direction is carry-forward, which worsens by 0.49 points at every budget because it never touches the calibration sample and so registers
the target change alone.

We report coverage against both targets throughout, and the full point-estimate
comparison under both definitions is in the reproduction package. The literature on
reconciling a weighted estimand with an unweighted likelihood, through pseudo-likelihood
and related devices, is reviewed by Parker et al. (2023); we do not use those methods
here, and the diagnostic above is what we offer in their place.
Whether weighting helps or hurts precision is itself dependent on how the
weighting variable relates to the outcome (Little \& Vartivarian, 2005); our choice
is specific to the small-sample cell-count setting and we do not generalise it.

\textbf{Supplementary Note S4. Deviations from the preregistration, in full.}

We list every departure from the registered plan.

\textbf{Budgets are nested rather than independent.} The registered randomisation procedure, followed to the letter, makes each budget's calibration sample a subset of the
next one's (Design). We reran the entire tournament drawing every condition's splits from an
independent random stream. Across the 118 registered contrast cells the mean paired difference moves by a
median of 0.04 points, and the verdict changes in two, both on the decision rule's
boundary and one of them against the synthetic panel. The between-budget reading
is stable: the direct estimator's learning curve is 57.7 $\times$ $n^{-0.367}$
registered against 58.2 $\times$ $n^{-0.369}$ independent, the budget MRP matches
at n = 100 is 384 against 380 in equivalent direct-estimation respondents, no
cost-equivalence multiplier moves by more than
0.1, and the (a) minus (d) interval excludes zero at all five budgets under both
runs. The registered run stands as the primary result and
the independent run is a deviation sensitivity analysis; both are in the repository.

\textbf{The registration's stated basis for 200 splits was wrong.} The registered text
argued that more splits shrink the percentile interval without limit and that at
200 splits a mean difference above roughly a seventh of the split standard
deviation would lift the 2.5th percentile above zero. That conflates the Monte
Carlo error of a percentile estimate with the width of the split-difference
distribution itself, which does not vanish; the one-seventh figure is 1.96/$\sqrt{200}$,
the confidence interval of the mean. We kept 200 splits and the registered rule
unchanged, since the rule as implemented reports percentiles of the split
distribution, and we correct the justification here rather than amend a fixed
public record.

\textbf{Contrast count for H4.} The registration sized H4 at 20 contrasts; we ran 40, because the same
registration forbids averaging (h) and (i) over the two generating models and
reporting each separately doubles the count. We followed the reporting rule and
record that the stated number did not carry it through. No contrast was added or
dropped after seeing results.

\textbf{Previous-wave source under temporal holdout.} The registration's estimator
table names the 2023 wave as the source for (f) and (g) without qualification,
while its baseline rule adds (f) to the temporal axis as an out-of-time baseline,
which implies the wave immediately before the target. The implementation follows
the baseline rule for (f), which uses the 2024 wave on that axis, and the table
for (g), whose prior stays at 2023 on every axis. Both choices were fixed in code
before the tournament ran and neither was changed after seeing results; a reader
who wants (g) on a 2024 prior should note that this would combine the full 2024
wave with a sample drawn from it.

\textbf{A design point outside the registered cell set.} After the tournament we found
that two of the fourteen registered cells lie outside the persona source's
population (Results) and recomputed the cell-holdout comparison without them, as a
post-hoc sensitivity analysis reported alongside the registered result rather than
in place of it.

\textbf{Convergence statistics computed after the run.} The registration requires
per-condition divergence counts, which we report, but not R-hat or effective
sample size; we computed those afterwards for a subset of conditions
(Implementation validation) and changed nothing as a result.

\textbf{Registered analyses not performed.} Three of the seven planned exploratory
analyses were not run: alternative auxiliary margins for raking, the shrinkage
parameter's trajectory in the budget, and the reversed temporal axis. The full
pairwise matrix is published for mean absolute error only; the ranking under all
three metrics is reported instead, with the one systematic contradiction described
in the Results.

\noindent\begin{minipage}{\textwidth}
\textbf{Supplementary Table S1. H4 under the registered fourteen-cell design and the twelve-cell sensitivity analysis.}

\nopagebreak[4]
\begin{center}\scriptsize\setlength{\tabcolsep}{3pt}
\begin{tabular}{@{}>{\raggedright\arraybackslash}p{0.037\textwidth}>{\raggedright\arraybackslash}p{0.106\textwidth}>{\raggedright\arraybackslash}p{0.071\textwidth}>{\raggedright\arraybackslash}p{0.114\textwidth}>{\raggedright\arraybackslash}p{0.128\textwidth}>{\raggedright\arraybackslash}p{0.106\textwidth}>{\raggedright\arraybackslash}p{0.071\textwidth}>{\raggedright\arraybackslash}p{0.114\textwidth}>{\raggedright\arraybackslash}p{0.128\textwidth}@{}}
\toprule
n & 14 cells: difference (MCSE) & 14 cells: interval & 14 cells: share favouring (h) & 14 cells: verdict & 12 cells: difference (MCSE) & 12 cells: interval & 12 cells: share favouring (h) & 12 cells: verdict \\
\midrule
50 & $-$1.56 (0.08) & [$-$3.79, 0.47] & 0.93 & Not distinguishable & $-$2.50 (0.08) & [$-$4.71, $-$0.59] & 0.99 & Synthetic leads \\ \addlinespace[2pt]
100 & 0.14 (0.06) & [$-$1.68, 1.68] & 0.40 & Not distinguishable & $-$0.40 (0.06) & [$-$2.21, 1.26] & 0.66 & Not distinguishable \\ \addlinespace[2pt]
200 & 1.55 (0.05) & [$-$0.11, 2.86] & 0.04 & Not distinguishable & 1.28 (0.06) & [$-$0.48, 2.50] & 0.07 & Not distinguishable \\ \addlinespace[2pt]
400 & 2.55 (0.05) & [1.25, 3.86] & 0.00 & MRP leads & 2.37 (0.05) & [1.16, 3.61] & 0.00 & MRP leads \\ \addlinespace[2pt]
800 & 3.30 (0.04) & [2.39, 4.41] & 0.00 & MRP leads & 3.13 (0.04) & [2.20, 4.22] & 0.00 & MRP leads \\ \addlinespace[2pt]
\bottomrule
\end{tabular}
\bigskip
\end{center}
\emph{Paired difference in cell MAE between the corrected synthetic estimator (h) and MRP (d) under cell holdout, in percentage points; negative favours the synthetic panel. Intervals are the 2.5th and 97.5th percentiles over 200 shared calibration draws. The fourteen-cell column is the registered primary analysis. The twelve-cell column excludes the two teen cells that lie outside the persona source's population, a post-hoc subset that favours the synthetic side; the verdict changes at n = 50 only. MCSE is the Monte Carlo standard error of the mean difference over the 200 draws; the share column is the proportion of draws on which (h) had the lower error. The three readings answer different questions, and the verdict follows the registered interval rule alone.}
\end{minipage}\par\bigskip

\noindent\begin{minipage}{\textwidth}
\textbf{Supplementary Table S2. Accuracy expressed as the budget direct estimation would need to match it.}

\nopagebreak[4]
\begin{center}\scriptsize\setlength{\tabcolsep}{3pt}
\begin{tabular}{@{}>{\raggedright\arraybackslash}p{0.293\textwidth}>{\raggedright\arraybackslash}p{0.052\textwidth}>{\raggedright\arraybackslash}p{0.060\textwidth}>{\raggedright\arraybackslash}p{0.060\textwidth}>{\raggedright\arraybackslash}p{0.060\textwidth}>{\raggedright\arraybackslash}p{0.060\textwidth}>{\raggedright\arraybackslash}p{0.198\textwidth}>{\raggedright\arraybackslash}p{0.103\textwidth}@{}}
\toprule
Estimator & n = 50 & n = 100 & n = 200 & n = 400 & n = 800 & Equivalent n at n = 100 & Basis \\
\midrule
(f) Carry-forward & 36.1 & 18.1 & 9.0 & 4.5 & 2.3 & 1,806 & Extrapolated \\ \addlinespace[2pt]
(g) Carry-forward + updating & 19.5 & 16.2 & 11.3 & 8.8 & 6.6 & 1,619 & Extrapolated \\ \addlinespace[2pt]
(d) MRP & 3.4 & 3.8 & 4.0 & 4.1 & 3.8 & 384 & Interpolated \\ \addlinespace[2pt]
(e) Beta-binomial & 2.2 & 2.6 & 3.0 & 3.1 & 3.0 & 256 & Interpolated \\ \addlinespace[2pt]
(h) Synthetic + correction, EXAONE & 3.5 & 2.2 & 1.3 & 0.8 & 0.4 & 218 & Interpolated \\ \addlinespace[2pt]
(h) Synthetic + correction, Gemini & 3.0 & 1.9 & 1.1 & 0.7 & 0.4 & 187 & Interpolated \\ \addlinespace[2pt]
(i) Synthetic prior, EXAONE & 2.2 & 2.3 & 2.4 & 2.4 & 2.5 & 233 & Interpolated \\ \addlinespace[2pt]
(i) Synthetic prior, Gemini & 2.5 & 2.1 & 1.9 & 1.9 & 2.1 & 210 & Interpolated \\ \addlinespace[2pt]
(c) Age-sex regression & 2.0 & 1.5 & 1.0 & 0.6 & 0.3 & 148 & Interpolated \\ \addlinespace[2pt]
(b) Raking & 1.2 & 0.9 & 0.9 & 0.9 & 1.1 & 93 & Interpolated \\ \addlinespace[2pt]
(j) Grand mean (floor) & 1.2 & 0.7 & 0.4 & 0.2 & 0.1 & 69 & Interpolated \\ \addlinespace[2pt]
\bottomrule
\end{tabular}
\bigskip
\end{center}
\emph{Each numbered column gives the multiple of the calibration budget that direct estimation would require to reach the same cell MAE, obtained by inverting the fitted learning curve MAE = 57.7 $\times$ $n^{-0.367}$ ($R^2$ = 0.982) for direct estimation without holdout. A multiple below one means the estimator is worse than simply buying that many more respondents. Values marked Extrapolated lie beyond the largest budget tested and are reported for order of magnitude only.}
\end{minipage}\par\bigskip

\noindent\begin{minipage}{\textwidth}
\textbf{Supplementary Table S3. Adjacent studies on the five points that fix what a comparison can show.}

\nopagebreak[4]
\begin{center}\footnotesize\setlength{\tabcolsep}{3pt}
\begin{tabular}{@{}>{\raggedright\arraybackslash}p{0.147\textwidth}>{\raggedright\arraybackslash}p{0.155\textwidth}>{\raggedright\arraybackslash}p{0.155\textwidth}>{\raggedright\arraybackslash}p{0.150\textwidth}>{\raggedright\arraybackslash}p{0.152\textwidth}>{\raggedright\arraybackslash}p{0.155\textwidth}@{}}
\toprule
Study & Purpose & Estimand & Budget variation & Unobserved segments & Baseline the synthetic or external source is judged against \\
\midrule
Kim and Cho (2026) & Validate a Korean persona panel against the same survey and test signature bias correction & Sex-by-age cell rates & Calibration fraction 1 to 30 per cent of the wave & One cell held out at a time & Direct estimation and age-sex regression, real data only \\ \addlinespace[2pt]
Wi\'sniowski et al. (2020) & Use a nonprobability sample as prior information for a probability-sample analysis & Regression coefficients and population means & Probability sample 50 to 1,000 & Not addressed & Probability-sample-only Bayesian regression \\ \addlinespace[2pt]
Huang et al. (2025) & Turn LLM-simulated responses into confidence sets for human population quantities & Population parameters, not by segment & Simulation sample size chosen adaptively & Not addressed & Human-only inference, expressed as human-respondent equivalents \\ \addlinespace[2pt]
Ye and Yoganarasimhan (2026) & Allocate a fixed budget of human respondents across questions when LLM predictions exist for all & Question-level population means, with extensions to regression coefficients & Budget fixed; efficiency reported at several budget levels & Not addressed; every question keeps human respondents & Uniform allocation with human-only means \\ \addlinespace[2pt]
Choi et al. (2026) & Align an LLM survey simulator to a small human pilot & Marginal, structural and individual fidelity of 36 Likert items & Pilot fixed at 5 per cent (n = 74) & Subgroup fidelity reported; no zero-pilot subgroup & Zero- and few-shot prompting, fine-tuning, rectification \\ \addlinespace[2pt]
Gordon et al. (2026) & Test whether richer persona context improves LLM survey simulation, and whether post-stratifying the synthetic data through the human MRP pipeline reduces residual error & National response distributions and plurality winners for 21 items & Human sample fixed at N = 996 & Not addressed; subgroup crosstab error discussed & Raw human distribution, human MRP estimates and a sampling-noise floor \\ \addlinespace[2pt]
This study & Race two synthetic-source constructions against small-area estimators under a preregistered rule & Sex-by-age cell rates & Calibration budget 50 to 800 & One cell held out at a time; two-cell subset as sensitivity & MRP, beta-binomial shrinkage, previous wave as prior and as carry-forward, direct estimation, raking, regression \\ \addlinespace[2pt]
\bottomrule
\end{tabular}
\bigskip
\end{center}
\emph{The table states what each study set out to do, from its own text, and is not a ranking. Purpose and design differ enough that none of the listed studies runs the comparison made here, and this study does not attempt theirs. Entries for Huang et al. (2025), Ye and Yoganarasimhan (2026), Choi et al. (2026) and Gordon et al. (2026) follow the versions cited in the reference list.}
\end{minipage}\par\bigskip

\noindent\begin{minipage}{\textwidth}
\textbf{Supplementary Table S4. The synthetic panel, as fixed before this study.}

\nopagebreak[4]
\begin{center}\footnotesize\setlength{\tabcolsep}{3pt}
\begin{tabular}{@{}>{\raggedright\arraybackslash}p{0.132\textwidth}>{\raggedright\arraybackslash}p{0.417\textwidth}>{\raggedright\arraybackslash}p{0.403\textwidth}@{}}
\toprule
 & 2024 panel & 2025 panel \\
\midrule
Generating models & gemini-3.5-flash (thinking = low, top\_p = 1.0; stated knowledge cutoff January 2025) and K-EXAONE-236B-A23B (LG AI Research; thinking disabled; cutoff undisclosed) & Same two models \\ \addlinespace[2pt]
Generation period & Mid-2026 & Mid-2026, after the 2025 survey wave \\ \addlinespace[2pt]
Persona source & Nemotron-Personas-Korea (about one million records, 26 fields) & Same \\ \addlinespace[2pt]
Per-cell allocation & clip(2 $\times$ cell size in the survey, 200, 600) personas per sex-by-age cell & Same rule \\ \addlinespace[2pt]
Valid responses & Gemini 8,168; EXAONE 8,165 (cell minimum 478, maximum 600) & Gemini 7,938; EXAONE 7,794 (cell minimum 349, maximum 600) \\ \addlinespace[2pt]
Prompt & First-person persona block including cultural background, hobbies and media orientation; items in the survey's original Korean wording; answer returned as JSON mapping item codes to option numbers; temperature 1.0 (main-analysis set) & Same \\ \addlinespace[2pt]
Items generated & Eight binary usage items and eight five-point constructs & Twelve binary items, the eight usage items among them \\ \addlinespace[2pt]
Used here & Cell-level rates of the eight binary items (no holdout and cell holdout) & Cell-level rates of the eight binary items (temporal axis) \\ \addlinespace[2pt]
Source & persona-validation-repro v1.0, doi:10.5281/zenodo.21397425, MIT licence & Same \\ \addlinespace[2pt]
\bottomrule
\end{tabular}
\bigskip
\end{center}
\emph{No responses were generated for this study; every analysis is conditional on this fixed panel. Model identifiers, generation period, allocation rule and prompt conditions are as documented by Kim and Cho (2026), Sections III-A to III-D; response counts are from the released files.}
\end{minipage}\par\bigskip

\noindent\begin{minipage}{\textwidth}
\textbf{Supplementary Table S5. The hybrid against the registered MRP.}

\nopagebreak[4]
\begin{center}\footnotesize
\resizebox{\ifdim\width>\textwidth\textwidth\else\width\fi}{!}{%
\begin{tabular}{@{}llll@{}}
\toprule
Regime & n & EXAONE hybrid $-$ (d) & Gemini hybrid $-$ (d) \\
\midrule
No holdout & 50 & $-$0.47 [$-$0.96, 0.03] & $-$1.67 [$-$3.09, $-$0.07]$\dagger$ \\
No holdout & 100 & $-$0.19 [$-$0.62, 0.14] & $-$0.95 [$-$1.81, 0.02] \\
No holdout & 200 & $-$0.05 [$-$0.29, 0.21] & $-$0.53 [$-$1.14, 0.06] \\
No holdout & 400 & $-$0.04 [$-$0.21, 0.16] & $-$0.31 [$-$0.79, 0.13] \\
No holdout & 800 & $-$0.05 [$-$0.20, 0.08] & $-$0.18 [$-$0.46, 0.05] \\
Temporal holdout & 50 & $-$0.52 [$-$1.03, $-$0.03]$\dagger$ & $-$1.69 [$-$2.92, $-$0.39]$\dagger$ \\
Temporal holdout & 100 & $-$0.19 [$-$0.63, 0.14] & $-$0.89 [$-$1.69, 0.06] \\
Temporal holdout & 200 & $-$0.01 [$-$0.30, 0.29] & $-$0.40 [$-$1.03, 0.21] \\
Temporal holdout & 400 & 0.06 [$-$0.13, 0.33] & $-$0.14 [$-$0.62, 0.35] \\
Temporal holdout & 800 & 0.07 [$-$0.07, 0.24] & 0.00 [$-$0.33, 0.30] \\
Cell holdout & 50 & $-$0.83 [$-$1.43, $-$0.26]$\dagger$ & $-$2.75 [$-$4.18, $-$1.50]$\dagger$ \\
Cell holdout & 100 & $-$0.50 [$-$1.02, $-$0.04]$\dagger$ & $-$2.09 [$-$3.27, $-$0.98]$\dagger$ \\
Cell holdout & 200 & $-$0.28 [$-$0.76, 0.11] & $-$1.48 [$-$2.42, $-$0.62]$\dagger$ \\
Cell holdout & 400 & $-$0.16 [$-$0.60, 0.22] & $-$1.10 [$-$1.91, $-$0.43]$\dagger$ \\
Cell holdout & 800 & $-$0.13 [$-$0.51, 0.13] & $-$0.93 [$-$1.53, $-$0.29]$\dagger$ \\
\bottomrule
\end{tabular}}
\bigskip
\end{center}
\emph{Paired difference in cell MAE on the primary six-item plane, in percentage points, mean and 2.5th to 97.5th percentiles over 200 shared calibration draws; negative favours the hybrid. $\dagger$ marks a split-stable contrast (interval excluding zero). The hybrid (d+s) is the registered MRP with the logit of the synthetic panel's cell rate as a covariate, coefficient prior Normal(0, 0.5), one fit per generating model. Every entry is exploratory. Across the 1,280 batched hybrid fits (2 models $\times$ 5 budgets $\times$ 16 fold groups $\times$ 8 items; 2,000 retained transitions each, 2,560,000 in all) 27 transitions diverged, all at n = 800: 13 in the EXAONE fits and 11 in the Gemini fits under cell holdout, and 3 in the Gemini fits without holdout. By item, 13 were in fits of the AI-service item and 10 of the short-form item, both outside the primary six-item plane, and 4 on the primary plane (3 in the metaverse item without holdout, 1 in the content-subscription item under cell holdout); the largest count in any single fit was 8, of 2,000. The per-fit counts for those three conditions are in the reproduction package.}
\end{minipage}\par\bigskip

\noindent\begin{minipage}{\textwidth}
\textbf{Supplementary Table S6. The hybrid against the strengthened MRP.}

\nopagebreak[4]
\begin{center}\footnotesize
\resizebox{\ifdim\width>\textwidth\textwidth\else\width\fi}{!}{%
\begin{tabular}{@{}llll@{}}
\toprule
Regime & n & EXAONE hybrid $-$ (d2) & Gemini hybrid $-$ (d2) \\
\midrule
No holdout & 50 & 0.27 [$-$0.29, 0.90] & $-$0.92 [$-$1.86, 0.23] \\
No holdout & 100 & 0.29 [$-$0.18, 0.71] & $-$0.47 [$-$1.09, 0.15] \\
No holdout & 200 & 0.22 [$-$0.11, 0.53] & $-$0.26 [$-$0.73, 0.21] \\
No holdout & 400 & 0.12 [$-$0.13, 0.40] & $-$0.15 [$-$0.48, 0.17] \\
No holdout & 800 & 0.04 [$-$0.17, 0.22] & $-$0.09 [$-$0.31, 0.16] \\
Temporal holdout & 50 & 0.14 [$-$0.37, 0.61] & $-$1.03 [$-$1.98, $-$0.04]$\dagger$ \\
Temporal holdout & 100 & 0.21 [$-$0.16, 0.56] & $-$0.49 [$-$1.09, 0.10] \\
Temporal holdout & 200 & 0.21 [$-$0.10, 0.52] & $-$0.18 [$-$0.61, 0.35] \\
Temporal holdout & 400 & 0.16 [$-$0.07, 0.39] & $-$0.04 [$-$0.38, 0.35] \\
Temporal holdout & 800 & 0.10 [$-$0.08, 0.26] & 0.03 [$-$0.23, 0.23] \\
Cell holdout & 50 & 0.40 [$-$0.23, 0.97] & $-$1.53 [$-$2.59, $-$0.59]$\dagger$ \\
Cell holdout & 100 & 0.52 [$-$0.10, 1.08] & $-$1.07 [$-$1.72, $-$0.39]$\dagger$ \\
Cell holdout & 200 & 0.46 [0.00, 0.97]$\dagger$ & $-$0.74 [$-$1.47, $-$0.08]$\dagger$ \\
Cell holdout & 400 & 0.36 [$-$0.10, 0.76] & $-$0.59 [$-$1.30, $-$0.07]$\dagger$ \\
Cell holdout & 800 & 0.27 [$-$0.10, 0.63] & $-$0.54 [$-$1.06, 0.01] \\
\bottomrule
\end{tabular}}
\bigskip
\end{center}
\emph{Paired difference in cell MAE on the primary six-item plane, in percentage points, mean and 2.5th to 97.5th percentiles over 200 shared calibration draws; negative favours the hybrid. $\dagger$ marks a split-stable contrast (interval excluding zero). The hybrid (d+s) is the registered MRP with the logit of the synthetic panel's cell rate as a covariate, coefficient prior Normal(0, 0.5), one fit per generating model. (d2) is the strengthened MRP; both sides of every contrast were built after the tournament, and every entry is exploratory.}
\end{minipage}\par\bigskip

\noindent\begin{minipage}{\textwidth}
\textbf{Supplementary Table S7. Cell MAE by calibration budget and holdout regime.}

\nopagebreak[4]
\begin{center}\scriptsize\setlength{\tabcolsep}{3pt}
\begin{tabular}{@{}>{\raggedright\arraybackslash}p{0.149\textwidth}>{\raggedright\arraybackslash}p{0.318\textwidth}>{\raggedright\arraybackslash}p{0.056\textwidth}>{\raggedright\arraybackslash}p{0.065\textwidth}>{\raggedright\arraybackslash}p{0.065\textwidth}>{\raggedright\arraybackslash}p{0.065\textwidth}>{\raggedright\arraybackslash}p{0.065\textwidth}>{\raggedright\arraybackslash}p{0.103\textwidth}@{}}
\toprule
Holdout & Estimator & n = 50 & n = 100 & n = 200 & n = 400 & n = 800 & Beats floor \\
\midrule
No holdout & (f) Carry-forward & 3.67 & 3.67 & 3.67 & 3.67 & 3.67 & Yes \\ \addlinespace[2pt]
No holdout & (g) Carry-forward + updating & 4.60 & 3.82 & 3.38 & 2.87 & 2.47 & Yes \\ \addlinespace[2pt]
No holdout & (d) MRP & 8.76 & 6.48 & 4.93 & 3.79 & 3.02 & Yes \\ \addlinespace[2pt]
No holdout & (e) Beta-binomial & 10.21 & 7.52 & 5.53 & 4.19 & 3.30 & Yes \\ \addlinespace[2pt]
No holdout & (h) Synthetic + correction, EXAONE & 8.66 & 7.98 & 7.41 & 6.97 & 6.76 & Yes \\ \addlinespace[2pt]
No holdout & (h) Synthetic + correction, Gemini & 9.17 & 8.45 & 7.85 & 7.32 & 7.06 & Yes \\ \addlinespace[2pt]
No holdout & (i) Synthetic prior, EXAONE & 10.27 & 7.79 & 6.01 & 4.62 & 3.56 & Yes \\ \addlinespace[2pt]
No holdout & (i) Synthetic prior, Gemini & 9.80 & 8.09 & 6.52 & 5.00 & 3.76 & Yes \\ \addlinespace[2pt]
No holdout & (c) Age-sex regression & 10.62 & 9.21 & 8.23 & 7.64 & 7.29 & Yes \\ \addlinespace[2pt]
No holdout & (b) Raking & 13.00 & 10.90 & 8.66 & 6.55 & 4.71 & Yes \\ \addlinespace[2pt]
No holdout & (a) Direct & 13.04 & 10.89 & 8.67 & 6.54 & 4.71 & Yes \\ \addlinespace[2pt]
No holdout & (j) Grand mean (floor) & 13.00 & 12.21 & 11.73 & 11.44 & 11.30 & (floor) \\ \addlinespace[2pt]
Cell holdout & (f) Carry-forward & 3.67 & 3.67 & 3.67 & 3.67 & 3.67 & Yes \\ \addlinespace[2pt]
Cell holdout & (g) Carry-forward + updating & 3.67 & 3.67 & 3.67 & 3.67 & 3.67 & Yes \\ \addlinespace[2pt]
Cell holdout & (d) MRP & 11.09 & 8.74 & 6.86 & 5.50 & 4.59 & Yes \\ \addlinespace[2pt]
Cell holdout & (e) Beta-binomial & 14.21 & 13.33 & 12.87 & 12.67 & 12.62 & \textbf{No} \\ \addlinespace[2pt]
Cell holdout & (h) Synthetic + correction, EXAONE & 9.52 & 8.88 & 8.40 & 8.05 & 7.88 & Yes \\ \addlinespace[2pt]
Cell holdout & (h) Synthetic + correction, Gemini & 10.08 & 9.39 & 8.91 & 8.46 & 8.23 & Yes \\ \addlinespace[2pt]
Cell holdout & (i) Synthetic prior, EXAONE & 14.42 & 14.42 & 14.42 & 14.42 & 14.43 & \textbf{No} \\ \addlinespace[2pt]
Cell holdout & (i) Synthetic prior, Gemini & 13.65 & 13.65 & 13.65 & 13.65 & 13.65 & \textbf{No} \\ \addlinespace[2pt]
Cell holdout & (c) Age-sex regression & 12.13 & 10.92 & 10.04 & 9.42 & 9.02 & Yes \\ \addlinespace[2pt]
Cell holdout & (b) Raking & 12.65 & 12.05 & 11.65 & 11.43 & 11.31 & Yes \\ \addlinespace[2pt]
Cell holdout & (a) Direct & 13.70 & 12.96 & 12.48 & 12.23 & 12.09 & \textbf{No} \\ \addlinespace[2pt]
Cell holdout & (j) Grand mean (floor) & 13.70 & 12.96 & 12.48 & 12.23 & 12.09 & (floor) \\ \addlinespace[2pt]
Temporal holdout & (f) Carry-forward & 2.55 & 2.55 & 2.55 & 2.55 & 2.55 & Yes \\ \addlinespace[2pt]
Temporal holdout & (g) Carry-forward + updating & 5.32 & 4.68 & 4.31 & 3.84 & 3.52 & Yes \\ \addlinespace[2pt]
Temporal holdout & (d) MRP & 9.74 & 7.35 & 5.75 & 4.53 & 3.83 & Yes \\ \addlinespace[2pt]
Temporal holdout & (e) Beta-binomial & 11.29 & 8.45 & 6.38 & 4.94 & 4.09 & Yes \\ \addlinespace[2pt]
Temporal holdout & (h) Synthetic + correction, EXAONE & 9.46 & 9.08 & 8.80 & 8.50 & 8.40 & Yes \\ \addlinespace[2pt]
Temporal holdout & (h) Synthetic + correction, Gemini & 11.59 & 11.13 & 10.73 & 10.36 & 10.19 & Yes \\ \addlinespace[2pt]
Temporal holdout & (i) Synthetic prior, EXAONE & 10.12 & 8.11 & 6.43 & 5.08 & 4.20 & Yes \\ \addlinespace[2pt]
Temporal holdout & (i) Synthetic prior, Gemini & 9.76 & 7.99 & 6.63 & 5.37 & 4.36 & Yes \\ \addlinespace[2pt]
Temporal holdout & (c) Age-sex regression & 11.04 & 9.84 & 9.08 & 8.54 & 8.30 & Yes \\ \addlinespace[2pt]
Temporal holdout & (b) Raking & 12.58 & 10.58 & 8.51 & 6.61 & 5.11 & Yes \\ \addlinespace[2pt]
Temporal holdout & (a) Direct & 12.61 & 10.56 & 8.51 & 6.60 & 5.11 & Yes \\ \addlinespace[2pt]
Temporal holdout & (j) Grand mean (floor) & 13.70 & 13.02 & 12.62 & 12.34 & 12.22 & (floor) \\ \addlinespace[2pt]
\bottomrule
\end{tabular}
\bigskip
\end{center}
\emph{Values are mean absolute error in percentage points on the primary six-item plane, averaged over 200 shared household-clustered calibration draws. Column headings give the calibration budget n. 'Beats floor' marks whether the estimator outperforms the uninformative grand mean (j) in that regime; estimators marked No fail there and are reported as failing rather than by rank alone. The n = 100 no-holdout column re-presents the pilot design point that informed the registered predictions. Under cell holdout (a) and (j) coincide exactly, and (f) and (g) to within 0.003 points, the residue of prior clipping.}
\end{minipage}\par\bigskip

\noindent\begin{minipage}{\textwidth}
\textbf{Supplementary Table S8. Updating versus correcting on unobserved segments.}

\nopagebreak[4]
\begin{center}\footnotesize\setlength{\tabcolsep}{3pt}
\begin{tabular}{@{}>{\raggedright\arraybackslash}p{0.323\textwidth}>{\raggedright\arraybackslash}p{0.162\textwidth}>{\raggedright\arraybackslash}p{0.291\textwidth}>{\raggedright\arraybackslash}p{0.075\textwidth}>{\raggedright\arraybackslash}p{0.075\textwidth}@{}}
\toprule
Estimator & External source & Attachment & n = 100 & n = 800 \\
\midrule
(f) Carry-forward, plain & Previous wave & None (source used directly) & 3.67 & 3.67 \\ \addlinespace[2pt]
(g) Carry-forward + updating & Previous wave & Bayesian updating & 3.67 & 3.67 \\ \addlinespace[2pt]
Carry-forward + correction & Previous wave & Bias correction & 5.60 & 3.27 \\ \addlinespace[2pt]
(h) Synthetic + correction & Synthetic panel & Bias correction & 8.88 & 7.88 \\ \addlinespace[2pt]
(i) Synthetic prior + updating & Synthetic panel & Bayesian updating & 14.42 & 14.43 \\ \addlinespace[2pt]
\bottomrule
\end{tabular}
\bigskip
\end{center}
\emph{Cell MAE in percentage points under cell holdout, primary six-item plane. Two external sources are each attached in two ways. Updating is flat in the budget for both sources, because an unobserved cell leaves the posterior mean at the prior mean. Correction transfers structure but must be estimated from the calibration sample: it pays at every budget for the heavily biased synthetic panel and costs more than it returns at n = 100 for the already accurate previous wave. The carry-forward analogue was built for this comparison and was not a registered estimator.}
\end{minipage}\par\bigskip
\par\endgroup

\end{document}